\documentclass[aip,jcp,amsmath,amssymb,reprint]{revtex4-1}

\usepackage{graphicx}
\usepackage{dcolumn}
\usepackage{bm}
\usepackage{mathrsfs}
\usepackage[dvipsnames]{xcolor}
\usepackage[normalem]{ulem}
\usepackage{caption}
\usepackage[textfont=rm]{subcaption}
\usepackage{appendix}

\begin{document}
 
\title{Thermodynamics and its correlation with dynamics in a mean-field model and pinned systems: A comparative study using two different methods of entropy calculation.}

\author{Ujjwal Kumar Nandi}
\thanks{\textit{Ujjwal Kumar Nandi and Palak Patel contributed equally to this work}}
\affiliation{\textit{Polymer Science and Engineering Division, CSIR-National Chemical Laboratory, Pune-411008, India}}
\affiliation{\textit{Academy of Scientific and Innovative Research (AcSIR), Ghaziabad 201002, India}}

\author{Palak Patel}
\thanks{\textit{Ujjwal Kumar Nandi and Palak Patel contributed equally to this work}}
\affiliation{\textit{Polymer Science and Engineering Division, CSIR-National Chemical Laboratory, Pune-411008, India}}
\affiliation{\textit{Academy of Scientific and Innovative Research (AcSIR), Ghaziabad 201002, India}}

\author{Mohd Moid}
\affiliation{\textit{Centre for Condensed Matter Theory, Department of Physics, Indian Institute of Science, Bangalore 560012, India}}

\author{Manoj Kumar Nandi}
\affiliation{\textit{Department of Engineering
University of Campania "Luigi Vanvitelli" 81031 Aversa (CE), Italy}}

\author{Shiladitya Sengupta}
\affiliation{\textit{Department of Physics, Indian Institute of Technology, Roorkee-247667, India}}

\author{Smarajit Karmakar}
\affiliation{\textit{Centre for Interdisciplinary Sciences, Tata Institute of Fundamental Research, 36/P, Gopanpally Village, Serilingampally Mandal, RR District, Hyderabad 500019, India}}

\author{Prabal K Maiti}
\affiliation{\textit{Centre for Condensed Matter Theory, Department of Physics, Indian Institute of Science, Bangalore 560012, India}}

\author{Chandan Dasgupta}
\affiliation{\textit{International Centre for Theoretical Sciences, Tata Institute of Fundamental Research, Bengaluru 560089, India}}
\affiliation{\textit{Centre for Condensed Matter Theory, Department of Physics, Indian Institute of Science, Bangalore 560012, India}}


\author{Sarika Maitra Bhattacharyya}
\email{mb.sarika@ncl.res.in}
\affiliation{\textit{Polymer Science and Engineering Division, CSIR-National Chemical Laboratory, Pune-411008, India}}
\affiliation{\textit{Academy of Scientific and Innovative Research (AcSIR), Ghaziabad 201002, India}}

\date{\today}

\begin{abstract}

Recently, some of us developed a novel model glass-forming liquid with $k$ extra interactions with pseudo neighbours to each liquid particle over and above the regular interactions with its neighbours. Analysis of the structure and dynamics of these systems showed that with an increase in $k$ the systems have more mean-field like properties. This work presents an extensive study of the thermodynamics of the above-mentioned model for several values of $k$ and its correlation with the dynamics. We surprisingly find that the usual thermodynamic integration (TI) method of calculating the entropy provides unphysical results for this model. It predicts the vanishing of configurational entropy at state points at which both the collective and the single-particle dynamics of the system show complete relaxation. We then employ a new method known as the two-phase thermodynamics (2PT) method to calculate the entropy. We find that with an increase in $k$ the difference in the entropy computed using the two methods (2PT and TI) increases. We also find that in the temperature range studied, the entropy calculated via the 2PT method satisfies the Adam-Gibbs (AG) relationship between the relaxation time and the configurational entropy, whereas the entropy calculated via the TI method shows a strong violation of the same. We then apply the 2PT method to calculate the entropy in another system where some fractions of particles are pinned randomly in their equilibrium positions. This system also shows a similar breakdown of the AG relationship as reported earlier. We show that the difference in entropy calculated via the 2PT and TI methods increases with an increase in pinning density. We also find that when the entropy is calculated using the 2PT method, the AG relationship between the dynamics and the entropy holds.

\end{abstract}

\maketitle

\section{Introduction:}
\label{sec_intro}
The slowing down of the dynamics in supercooled liquids and its correlation with the thermodynamics of the system have been topics of intense research. There are several characteristic temperatures where both the thermodynamic and dynamic properties of the system change in a significant manner. At the onset temperature ($T_{onset}$), the relaxation dynamics of the system start to differ from that of a typical liquid because due to the lowering of temperature, the system begins to explore the underlying free energy landscape \cite{sastry_nature_1998}. This onset temperature can also be identified as the temperature where the pair part of the excess entropy becomes less than the total excess entropy of the system \cite{atreyee_onset,Palak_polydisperse_onset}.
Below $T_{onset}$, the temperature dependence of the dynamics can be described reasonably well by the so-called mode-coupling theory (MCT), which predicts a power-law divergence of the relaxation times at a dynamic transition temperature $T_c$.\cite{Gotze_MCT_1999} However, experimental and numerical studies found \cite{Du_cummins_knauss_light-scattering_1994,lunkenheimer_dynamics_in_CKN_1997,Kim_multi_t_correlation_2013,kob-andersen,Flenner_MCT_brownian_dynamic_2005,szamel-pre} that the relaxation time does not diverge at $T_c$ as predicted by the MCT, but instead shows a smooth crossover to weaker temperature dependence.
This crossover scenario is consistent with the predictions of the so-called random first-order transition (RFOT) theory  \cite{wolynes_lubchenko,kirk_woly1} and it has been related to the properties of the underlying potential energy landscape \cite{andrea_supercooled_liq_2009}. 

According to the RFOT theory and the phenomenological Adam-Gibbs (AG) theory \cite{Adam-Gibbs}, the low-temperature dynamics of supercooled liquid is controlled by its configurational entropy ($S_{c}$), which measures the number of possible distinct states accessible to the system. The AG theory predicts the following relationship between the $\alpha$ relaxation time ($\tau$) and the configurational entropy ($S_{c}$): $\tau=\tau_{0} \exp(-A/TS_{c})$ where $\tau_{0}$ is a microscopic timescale and $A$ is a system-dependent constant. Thus according to the AG theory, the temperature $T_{0}$ where the relaxation time diverges is the same as the Kauzmann temperature $T_{K}$ where the configurational entropy goes to zero \cite{kauzmann}. For a large number of systems the AG relationship is found to hold \cite{Adam-Gibbs,Berthier_AG_hold,adam-gibbs_hold1,Adam-Gibbs_hold2,Adam-Gibbs_hold3,adam-Gibbs_hold4,adam-gibbs_hold5,adam-gibbs_hold6,adam-gibbs_hold7,adam-gibbs_hold8,adam-gibbs_hold9}. There has been a recent study which showed that it is the diffusion coefficient which follows the AG relationship for the widest temperature range \cite{sastry_anshul_AG_relation}.

The validity of the AG theory in the form presented above has recently been challenged \cite{Berthier_AG_hold}. It has been argued that according to the RFOT theory, the reduction in the configurational entropy is related to the growth of a static correlation length over which the activation happens, giving rise to the relaxation process. This theory predicts a generalized AG relation given by  $\tau=\tau_{0} \exp(-A/TS^{\alpha}_{c})$, where $\alpha$ can be different from unity. It was further shown that the generalized AG relation holds \cite{Berthier_AG_hold} both in experiments and in simulations. Note that even according to the generalized AG relationship, the relaxation timescale should diverge below $T=T_{K}$ when the configurational entropy vanishes. 

In a recent study, some of us have developed a novel of glass-forming liquid where we can switch between a 3-dimensional liquid and a fully connected mean-field system in a continuous manner \cite{ujjwal_mf}. The parameter that is introduced to achieve this is $k$ added pseudo neighbours for each particle. The structure, dynamics, and dynamical heterogeneity of this model have been studied as a function of $k$. It was shown that the structure given by the radial distribution function (rdf) of the usual neighbours remains almost unchanged with $k$. However, the pseudo neighbours do contribute to the total rdf that shows a weaker modulation with distance, a typical mean-field like behaviour \cite{ujjwal_mf, mari-kurchan}. With increase in $k$, the dynamics also slows down and the transition temperatures ($T_{0},T_{c},T_{onset}$) move to higher values. The range over which a system follows the MCT power-law behaviour becomes wider with an increase in $k$. The heterogeneity decreases with an increase in $k$. Thus it was shown that with an increase in $k$ the system becomes more mean-field like.

The goal of the present work is to study the thermodynamic properties of this system and its correlation with the dynamics. In order to do so, we employ the well-known thermodynamic integration (TI) method to calculate the total entropy and the configurational entropy of the system \cite{sastry-onset}. We find that with an increase in $k$, the Kauzmann temperature becomes higher which is similar to that found for $T_{0}$. However, we also find a violation of the AG relation. As discussed before, the breakdown of the AG relationship is a possibility but for larger $k$ systems we find that the configurational entropy vanishes at temperatures close to the onset temperature where the dynamics is reasonably fast. In our opinion, this is an unphysical result not even supported by the generalized AG relationship. This implies that the TI method of entropy calculation needs to be re-examined.
We discuss the possible failure points of the TI method however at present we do not have the know how to incorporate the corrections.  

We thus employ a completely different method to calculate the entropy of the system, namely the two-phase thermodynamics (2PT) method. It is a well-known method \cite{lin2003two,lin2010two} that has provided accurate entropy values over a wide range of thermodynamic state points for the LJ fluid and different water models \cite{lin2003two,moid2021}.
We first test this model for a regular Kob-Anderson model system which is the $k=0$ system in the mean-field model. 
We compare the entropy values obtained via the TI and the 2PT methods and find them to be close to each other. We then employ the 2PT method for different mean-field systems and compare the results with those obtained by the TI method. We find that with an increase in $k$ the difference in entropy obtained by the two methods increases. We also find that using the entropy calculated via the 2PT method, the AG relationship holds in the range of temperature studied here. 

Similar to the mean-field system, there has been some discussion of the dynamics not following the entropy and the break down of the AG relationship when the entropy was calculated using the TI method in another model, namely randomly pinned systems \cite{walter_original_pinning,reply_by_kob,smarajit_chandan_dasgupta_original_pinning,reply_by_chandan_dasgupata}. Given the success of the 2PT method in describing the entropy of the mean-field system, we further employ it to calculate the entropy of the pinned system. We find that with the increase in the pinning density, the difference in entropy computed by the TI and the 2PT methods increases. We also show that in the temperature range studied, the pinned systems follow the AG relationship when the entropy is calculated via the 2PT method.

The rest of the paper is organized as follows: The system and simulation details are described in Sec.\ref{sec_details}. In Sec.\ref{sec_entropy}, we describe different methods for the calculation of entropy. In Sec.\ref{sec_mf} and Sec.\ref{sec_pinned}, we present the results of our analysis for the mean-field and pinned systems, respectively. We discuss the implication of the results in Sec.\ref{sec_discuss} and conclude in Sec.\ref{sec_conclusion}.

\section{Details of system and simulations}
\label{sec_details}
We have studied two different families of models. One is a mean-field system and the other is a pinned system. For both the systems we have used atomistic models which are simulated with two-component mixtures of classical particles (larger ``A'' and smaller ``B'' type), where particles of type {\it i} interact with those of type {\it j} with pair potential, $u(r_{ij})$,  where $r$ is the distance between the pair. 
$u(r_{ij})$ is described by a shifted and truncated Lennard-Jones (LJ) potential, as given by:
\begin{equation}
 u(r_{ij}) =
\begin{cases}
 u^{(LJ)}(r_{ij};\sigma_{ij},\epsilon_{ij})- u^{(LJ)}(r^{(c)}_{ij};\sigma_{ij},\epsilon_{ij}),    & r\leq r^{(c)}_{ij}\\
   0,                                                                                       & r> r^{(c)}_{ij}
\end{cases}
\label{ka_model}
\end{equation}
\noindent where $u^{(LJ)}(r_{ij};\sigma_{ij},\epsilon_{ij})=4\epsilon_{ij}[(\frac{\sigma_{ij}}{r_{ij}})^{12}-(\frac{\sigma_{ij}}{r_{ij}})^{6}]$ and
 $r^{(c)}_{ij}=2.5\sigma_{ij}$. 
We have used the Kob-Andersen model\cite{kob-andersen} and performed constant volume and constant temperature (Noose-Hoover thermostat and velocity rescaling) simulation (NVT). We use $\sigma_{AA}$ and $\epsilon_{AA}$ as the units of length and energy, setting the Boltzmann constant $k_B=1$. We have used reduced time unit in terms of $\sqrt{\frac{m_{A}\sigma_{AA}^{2}}{\epsilon_{AA}}}$ and masses of both types of particles are taken to be the same ($m_{A}=m_{B}$, set equal to unity). We have used 80\% of A particles and 20\% of B particles with the diameter $\sigma_{AA}$=1.0, $\sigma_{AB}$=0.8 and $\sigma_{BB}$=0.88. The interaction strengths between the particles are $\epsilon_{AA}$=1.0, $\epsilon_{AB}$=1.5 and $\epsilon_{BB}$=0.5.

\subsection{Mean-Field System}

The mean-field system is given by $N$ particles that interact with each other via a standard short-range potential. In addition, each particle interacts also with ``pseudo neighbors'', i.e.~particles that are not necessarily close in space.
Hence, the total interaction potential of the system is given by,

\begin{eqnarray}
U_{\rm tot}(r_{1},..r_{N})&=&\sum_{i=1}^{N}\sum_{j>i}^{N}u(r_{ij})+\frac{1}{2}\sum_{i=1}^{N}\sum_{j=1}^{k}u^{\rm pseudo}(r_{ij}) \; \;
\label{eq1}\\
&=&U+U^{\rm pseudo}_{k} \qquad .
\label{eq2}
\end{eqnarray}
\noindent

The first term on the right-hand side is the regular interaction between particles while the second term is the interaction each particle has with its pseudo neighbours. Here we consider the case in which the regular interaction is described by the Eq.\ref{ka_model}.

The interaction potential with the pseudo neighbours is modelled in terms of a modified shifted and truncated LJ potential,

\begin{eqnarray}
u^{\rm pseudo}(r_{ij})&=&u(r_{ij}-L_{ij}) \\
&=&4\epsilon_{ij}\Big[\Big(\frac{\sigma_{ij}}{r_{ij}-L_{ij}}\Big)^{12}-\Big(\frac{\sigma_{ij}}{r_{ij}-L_{ij}}\Big)^6\Big] \quad 
\end{eqnarray}
\noindent where $L_{ij}$ is a random variable defined below. In our simulations we impose the restriction that any two particles interact either via $u(r_{ij})$ or via $u^{\rm pseudo}(r_{ij})$. This condition determines how the pseudo neighbors and the values $L_{ij}$ are chosen for a given configuration equilibrated with the potential $u$: for each particle $i$, we select $k$ random numbers $L_{ij}$ in the range $r_c \leq L_{ij} \leq L_{\rm max}$, where $L_{\rm max} \leq L_{\rm box}/2-r_c$, with $L_{\rm box}$ the size of the simulation box. (The distribution of these random variables will be denoted by $\mathscr{P}(L_{ij})$ and in the following, we will consider the case that the distribution is uniform.) Subsequently we choose $k$ distinct  particles $j$ with $r_{ij}>r_c$ and use the $L_{ij}$ to fix permanently the interaction between particles $i$ and $j$. This procedure thus makes sure that each particle $i$ interacts not only with the particles that are within the cutoff distance but in addition to $k$ particles that can be far away. Note that once the particle $j$ is chosen as a pseudo neighbour of particle $i$, automatically particle $i$ becomes a pseudo neighbour of particle $j$. The system, as defined here, can then be simulated using standard simulation algorithms.

NVT molecular dynamics (MD) simulation is performed in a cubic box using velocity rescaling method for $N=2744$ particles at $\rho=1.2$ ( $L_{\rm box}=13.1745$), using a time integration step of $\Delta t=0.005$. For $L_{\rm max}$ we have taken 4.0, slightly below the maximum value of 4.09. We have simulated four different systems with the number of pseudo neighbours, $k=0,4,12,$ and 28.

\subsection{Pinned System}
\label{sec_details_pin}
For the study of the pinned system, we use the Kob-Andersen 80:20 binary Lenard-Jones mixture\cite{kob-andersen}. Details of this model are given in Sec.\ref{sec_details}. For creating the pinned system the following pinning protocol is used. The pinned particles are chosen randomly from an equilibrium configuration of the system at the temperature of interest \cite{walter_original_pinning,smarajit_chandan_dasgupta_original_pinning}. NVT molecular dynamics simulation is performed in a cubic box using Nose-Hoover thermostat where N=1000 at $\rho=1.2$ ($L_{box}$=9.41036) using a time integration step of $\Delta$t = 0.005, at three different pinning concentration ($c$), i.e. 0.05, 0.10 and 0.15. Production runs of pinned configurations are long enough to ensure that within the simulation time, the overlap correlation function Q(t) (defined in Sec.\ref{Dynamics_calculation}) decays to zero.

\subsection{Dynamics}
\label{Dynamics_calculation}
To analyze the dynamics, we consider the self part of the overlap function,

\begin{equation}
Q(t) =\frac{1}{N} \Big \langle \sum_{i=1}^{N}  \omega (|{\bf{r}}_i(t)-{\bf{r}}_i(0)|)\Big \rangle \quad 
\label{eq_self_overlap}
\end{equation}
\noindent where the function $\omega(x)$ is 1 if $0\leq x\leq a$ and $\omega(x)=0$ otherwise. The parameter $a$ is chosen to be 0.3, a value that is slightly larger than the size of the ``cage'' determined from the height of the plateau in the mean square displacement at intermediate times~\cite{kob-andersen}. Thus the quantity $Q(t)$ measures whether or not at time $t$ a tagged particle is still inside the cage it occupied at $t=0$. 

To analyze the collective dynamics of the systems, we have used both the collective overlap function and the collective intermediate scattering function. 

The collective overlap function is defined as follows,
\begin{equation}
Q^{tot}(t) =\frac{1}{N} \Big \langle \sum_{i=1}^{N}\sum_{j=1}^{N} \omega (|{\bf{r}}_i(t)-{\bf{r}}_j(0)|) \Big \rangle \quad ,
\label{eq_tot_overlap}
\end{equation}
\noindent
The long time saturation value of Q$^{tot}$(t) is given by (using a=0.3 ),\cite{shiladitya_2011}
\begin{equation}
\lim_{t \to \infty}Q^{tot}(t)=\frac{N}{V} \frac{4}{3}\pi a^{3}=0.135
\label{eq_collective_overlap}
\end{equation}
\noindent

We have also calculated the intermediate scattering function $F(q,t)$. It is the collective density-density time correlation function in momentum space which provides information about the collective dynamics of the systems.

\begin{equation}
F(q,t) = \frac{1}{N F(q,0)} \Big< \sum_{i=1}^{N} \sum_{j=1}^{N} \exp[-i\bf{q}.(\bf{r}_{i}(t)-\bf{r}_{j}(0))] \Big>  
\label{eq_fqt}
\end{equation}
\noindent

The relaxation time ($\tau$) is calculated from the self part of the overlap function, when it decays to 1/e . The rapid increase in relaxation time with decreasing temperature is a signature of glassy dynamics. This is often fitted to the Vogel-Fulcher-Tammann (VFT) equation.
\begin{equation}
\tau(T) = \tau_{0}\exp\Big[\frac{1}{K(\frac{T}{T_0}-1)}\Big] \quad .
\label{eq24}
\end{equation}
\noindent
Here $\tau_0$ is a high-temperature relaxation time and $T_0$ is the so-called VFT temperature at which the relaxation time of the system is predicted to diverge. The parameter $K$ describes the curvature of the data in an Arrhenius plot and hence can be considered as a measure for the fragility of the glass-former.

\section{Entropy}
\label{sec_entropy}
In this work, we have used two different well-known methods for the calculation of the total entropy ($S_{tot}$) of the system. Below we provide  brief sketches of the two methods, namely the TI method \cite{sastry-onset} 
and the 2PT method \cite{lin2003two} 

\subsection{Thermodynamic integration (TI) method}
Below we describe the different quantities required to calculate the entropy in the TI method \cite{sastry-onset} 

\subsubsection{Ideal gas entropy}
\label{ideal_gas}
Ideal gas entropy is the entropy of a set of non-interacting particles. The ideal gas entropy per particle for a binary system at temperature $T$ is given by 

\begin{equation}
S_{ideal} = \frac{5}{2}-\ln (\rho) + \frac{3}{2}\ln \Big(\frac{2\pi T}{h^2}\Big) + \frac{1}{N}\ln \frac{N!}{N_{A}!N_{B}!} 
\label{S_ideal_eq}
\end{equation}
\noindent where $N=N_{A}+N_{B}$ is the total number of particles, $V$ is the volume of the system and h is the Planck constant. $N_{A}$ and $N_{B}$ are number of particles of type A and B. The last term contributes to the mixing entropy.

However, if the particles are divided into 'M' distinguishable species such that $N = \sum_{i=1}^{M}N_{i}$ then the ideal gas entropy per particle can written as,
\begin{equation}
S_{ideal}^{d} = \frac{5}{2}-\ln (\rho) + \frac{3}{2}\ln \Big(\frac{2\pi T}{h^2}\Big) + \frac{1}{N} \ln \frac{N!}{\Pi_{i=1}^{M} N_{i}!}
\label{S_ideal_eq_disting}
\end{equation}

\noindent

\subsubsection{Excess entropy and Total entropy}
Excess entropy ($S_{ex}$) estimates the loss of entropy due to interactions among the particles. It is always a negative quantity. $S_{ex}$ is calculated using the TI method where the integration can be done on the temperature path \cite{walter_original_pinning}, in the temperature range $\infty$ to a target temperature ($T^{*}$). 

\begin{equation}
S_{ex}(\beta^{*})= \beta^{*} \big<U\big> -\int_{0}^{\beta^{*}}d\beta \big<U\big>
\label{S_ex_eq}
\end{equation}
\noindent 
Here $\beta=\frac{1}{T}$.
The total entropy of the system at a particular temperature is the sum of the ideal gas entropy and the excess entropy of the system at that particular temperature.
\begin{equation}
S_{tot}=S_{ideal}+S_{ex}
\label{total_TI}
\end{equation}
\noindent

\subsection{Two-phase thermodynamics (2PT) method}
The 2PT is another conventional method to compute the entropy of liquids \cite{lin2003two,lin2010two}. In the 2PT method, the thermodynamics quantities can be computed using the density of state (DOS) of the liquid. One can decompose the DOS of a liquid as a sum of solid-like and gas-like contributions. To compute the thermodynamic quantities, the phonons in the solid-like DOS are treated as non-interacting harmonic oscillators, as in the Debye model\cite{mcquarrie1976statistical}. On the other hand, the gas-like DOS is described as a low-density hard-sphere fluid, which can be computed analytically\cite{mcquarrie1976statistical}. Using the 2PT description, Lin {\it et al.},\cite{lin2003two,lin2010two} demonstrated that the thermodynamics quantities of the LJ fluid can be computed very accurately over a wide range of thermodynamics state points using a very small MD trajectory. In a later work, Lin {\it et al.} \cite{lin2012two} calculated the entropy of a binary fluid using the 2PT method. Here, we provide a brief overview of the decomposition of the DOS in the 2PT method. We refer the reader to the original papers\cite{lin2003two,lin2010two} for a full description.

The density of state function, g($\nu$), can be computed from the mass-weighted atomic spectral densities, defined as\cite{lin2003two,lin2010two},

\begin{equation}\label{eq:2pt4}
\text{g}(\nu) = \frac{2}{T} \sum_{j=1}^{N}\sum_{l=1}^{3} {m_js_j^l(\nu)}
\end{equation}
\noindent where $m_j$ is the mass of the $j^{th}$ atom, $l$ denotes the direction in the Cartesian coordinates, and $s_j^l(\nu)$ are the atomic spectral densities defined as,

\begin{equation}\label{eq:2pt5}
s_j^l(\nu) = \lim_{\tau\rightarrow\infty}\frac{\left|\int_{-\tau}^{\tau}v_j^l(t)e^{-i2\pi\nu t}dt\right|^2}{2\tau}\,
\end{equation}
\noindent where $v_j^l(t)$ denotes the velocity component of the $j^{th}$ atom in the $l^{th}$ direction. The atomic spectral density, $s_j^l(\nu)$, can be computed from the Fourier transform of the velocity auto-correlation function (VACF) $c_j^l(t)$.

\begin{equation}\label{eq:2pt6}
s_j^l(\nu) = \lim_{\tau\rightarrow\infty} \int_{-\tau}^{\tau}{c_j^l(t)e^{-i2\pi\nu t}dt}
\end{equation}
\noindent where $c_{j}^{l}(t)$ is given by:
\begin{equation}\label{eq:2pt7}
c_j^l(t) = \lim_{\tau\rightarrow\infty} \frac{1}{2\tau} \int_{-\tau}^{\tau}{v_j^l(t+t^\prime)v_j^l(t^\prime)dt^\prime}
\end{equation}
\noindent
Thus, Eq.\ref{eq:2pt4} can be rewritten as:

\begin{equation}\label{eq:2pt8}
\text{g}(\nu) = \frac{2}{T} \lim_{\tau\rightarrow\infty} \int_{-\tau}^{\tau} {\sum_{j=1}^{N}\sum_{l=1}^{3} {m_jc_j^l(t)e^{-i2 \pi \nu t}dt} }
\end{equation}
\noindent

As we mentioned above, g($\nu$) can be decomposed into solid and gas-like components in the 2PT formalism. Based on the diffusivity of the system compared to hard-sphere gas at the same density, Lin {\it et al.}\cite{lin2003two} proposed a self-consistent fluidity factor, $f$, which decides the degrees of freedom shared in solid and gas components. The relationship between $f$ and dimensionless diffusivity, $\Delta$, can be derived (for the details of the derivation, readers are referred to ref \cite{lin2003two}).

\begin{equation}
\begin{split}\label{eq:2pt12}
2\Delta^{-9/2}f^{15/2}-6\Delta^{-3}f^{5}-\Delta^{-3/2}f^{7/2}+ \\ 6\Delta^{-3/2}f^{5/2}+2f-2=0
\end{split}
\end{equation}
\noindent

The dimensionless diffusivity constant, $\Delta$, depends on the material properties.

\begin{equation}\label{eq:2pt13}
\Delta(T,\rho,m,\text{g}_0) = \frac{2\text{g}_0}{9N}\left(\frac{6}{\pi}\right)^{2/3}\left(\frac{\pi T}{m}\right)^{1/2}\rho^{1/3}
\end{equation}
\noindent where, g$_0$ = g(0) is the DOS of the system at zero-frequency and $\rho$ is the number density.  Using $f$ obtained from
Eqs.\ref{eq:2pt12}, \ref{eq:2pt13}, the DOS in the gas-like diffusive component can be obtained using a
hard-sphere diffusive model:

\begin{equation}\label{eq:2pt14}
\text{g}^g(\nu) = \frac{g_0}{1+\left[\frac{\pi g_0\nu}{6fN}\right]^2}
\end{equation}
\noindent

Given the DOS in the gas-like component, one can compute the solid-like DOS, g$^s(\nu$),  using the equation

\begin{equation}\label{eq:2pt15}
    \text{g}(\nu) = \text{g}^\text{g}(\nu)  + \text{g}^s(\nu)
\end{equation}
\noindent

Once the decomposition of the DOS has been done, any thermodynamic quantity, $A$, can be computed using the corresponding weight functions,

\begin{equation}\label{eq:2pt16}
A = \beta^{-1}\left[\int_{0}^{\infty}{ \text{g}^g(\nu)W_{A}^{g}}d\nu + \int_{0}^{\infty}{ \text{g}^s(\nu)W_{A}^{s}}d\nu\right]
\end{equation}
\noindent

The weight functions for the entropy in the solid (W$_{S}^s$) and the gas-like component (W$_{S}^g$) component are defined as:
\begin{equation}\label{eq:2pt17}
\text{W$_{S}^{s}$}(\nu) = \text{W$_S^{HO}$}(\nu) =\frac{\beta\hbar\nu}{\exp{(\beta\hbar\nu)} - 1} -\ln{[1 - \exp{(-\beta\hbar\nu)}]} 
\end{equation}
\noindent where $\beta$ = $\frac{1}{T}$ and $\hbar$ = $\frac{h}{2\pi}$, h is Planck constant.
\begin{equation}\label{eq:2pt18}
\text{W$_{S}^{g}$}(\nu) =\frac{1}{3} \frac{S^{HS}}{k}
\end{equation}
\noindent where, $S^{HS}$ denotes the entropy of the hard sphere system. Using Eqs.\ref{eq:2pt17}, \ref{eq:2pt18}, the total entropy of the system can be written as,

\begin{equation}
  S_{tot} = S^s + S^\text{g}
\label{total_2PT}
\end{equation}
\noindent

In this work, for the calculation of the entropy using the 2PT method, we have averaged over ten data sets where each data set starts with a different configuration and velocity distribution. Each data set contains fifty thousand frames of velocity with an interval of 0.005 time steps.

\subsection{Configurational entropy}

As discussed earlier we can calculate the total entropy using both the TI and the 2PT methods. Thus Eqs.\ref{total_TI} and \ref{total_2PT} provide us with the same information although the routes of obtaining them are different. 

In the supercooled liquid regime, the configurational space can be divided into inherent structure minima and vibrational motion around them. The logarithm of the number of these inherent structure minima gives the configurational entropy ($S_{c}$) of the system, which can be calculated by subtracting the vibrational entropy, $S_{vib}$ from the total entropy of the system.

\begin{equation}
\begin{aligned}
 S_{c} &= S_{tot} - S_{vib} \\
 &= S_{ideal} + S_{ex} - S_{vib} 
\end{aligned}
\label{S_c_eq}
\end{equation}
\noindent

The vibrational entropy is calculated by making a harmonic approximation about a local minimum \cite{sastry-nature,shiladitya_2011,Sciortino_2005,Heuer_2008}.  To obtain the vibrational frequencies we calculate the Hessian and then diagonalize it. Once we obtain the vibrational frequencies, $S_{vib}$ is calculated using the following equation,

\begin{equation}
S_{vib} = \frac{3}{2}\ln \Big(\frac{2\pi T}{h^{2}} \Big)+ \frac{\ln(V)}{N}+\frac{1}{2N}\sum_{i=1}^{3N-3}\ln \Big(\frac{2\pi T}{{\omega_{i}}^{2}} \Big)-\frac{3}{2N} + 3
\label{S_vib_eq}
\end{equation}
\noindent

\section{Results for Mean-field system}
\label{sec_mf}

In this section, we will discuss the entropy of the mean-field system and its correlation with the dynamics. We will first discuss the results obtained using the TI method and its shortcomings and then discuss the results obtained from the 2PT method.   

\subsection{Entropy using thermodynamic integration method}
In the estimation of the entropy using the TI method, we need to calculate the excess entropy and the vibrational entropy. The configurational entropy is then obtained from Eq.\ref{S_c_eq}.

\subsubsection{Excess entropy}
Note that in the calculation of the excess entropy via the TI method, we need the information of the internal energy (Eq.\ref{S_ex_eq}). For the mean-field systems, the internal energy has two parts, one is the contribution from the regular neighbour (NN) and the other is the contribution from the pseudo-neighbour (PN).
A similar decomposition is present for the entropy, where we can write $S_{ex}=S_{ex}^{NN}+S_{ex}^{PN}$. The first term on the r.h.s refers to the contribution from the regular neighbours and the second term from that of the pseudo neighbours. These are given by,

\begin{eqnarray}
S_{ex}^{NN}(\beta^{*},k)= \beta^{*} \big<U\big> -\int_{0}^{\beta^{*}}d\beta \big<U\big>
\label{S_ex_eq_mf_NN}
\end{eqnarray}
\noindent
and
\begin{eqnarray}
 S_{ex}^{PN}(\beta^{*},k)=\beta^{*} \big<U_{k}^{pseudo}\big> -\int_{0}^{\beta^{*}}d\beta \big<U_{k}^{pseudo}\big>\Big]
\label{S_ex_eq_mf_PN}
\end{eqnarray}
\noindent

\begin{figure}[!bth]
\begin{subfigure}[h]{0.4\textwidth}
\includegraphics[width=0.9\textwidth,trim = {0 0cm 0 0.0cm},clip]{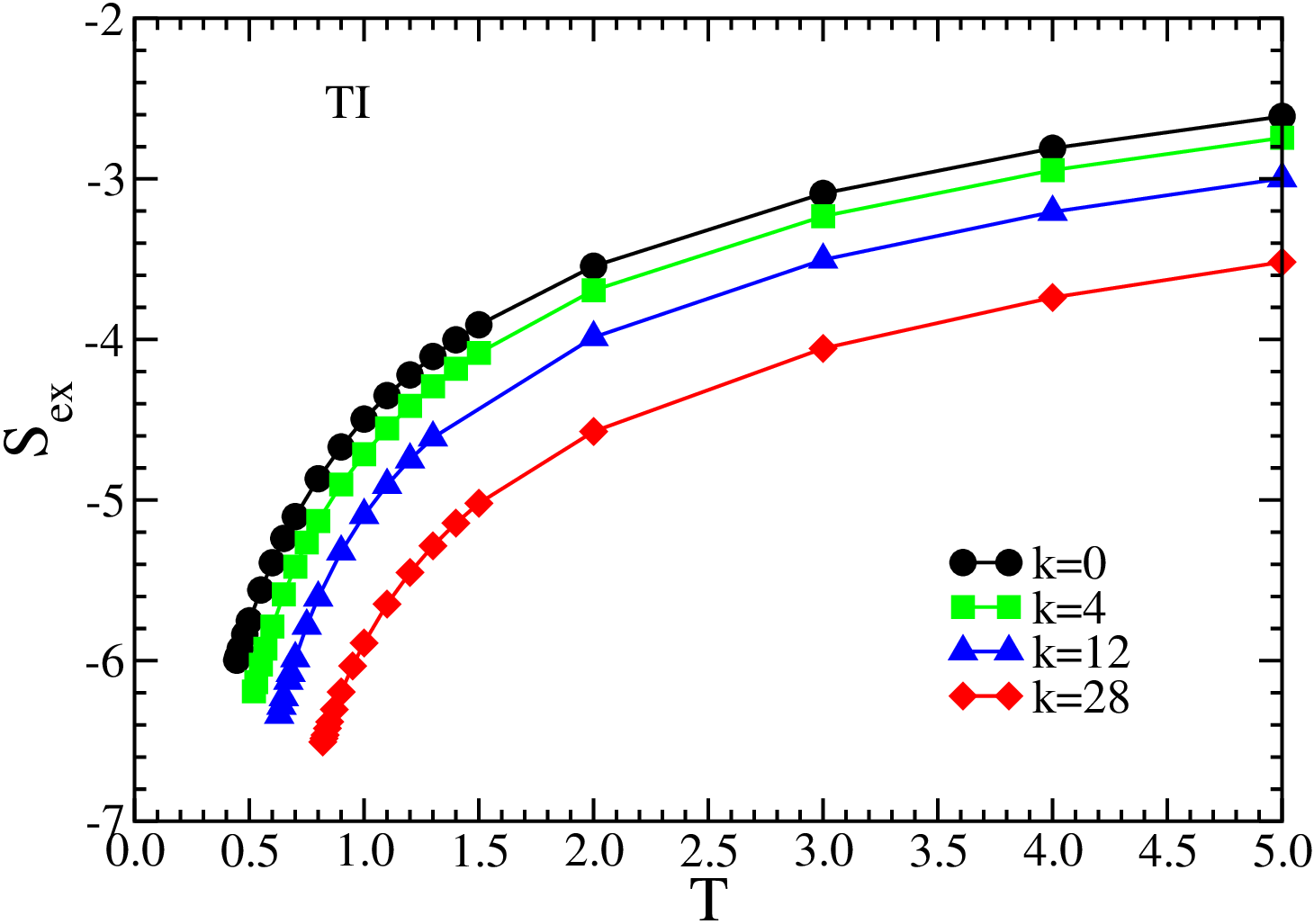}
\end{subfigure}
\caption{\emph {Plot of per particle excess entropy $S_{ex} \, vs. \, T$ for $k=0,4,12$ and 28 systems. $S_{ex}$ is estimated using the TI method. With increase in $k$, the excess entropy becomes more negative.}}
\label{s_ex_all_k}
\end{figure}

In Fig.\ref{s_ex_all_k}, we plot the temperature dependence of $S_{ex}$ from the TI method for different $k$ systems. In the TI method, we assume the particles to be indistinguishable. We find that the excess entropy decreases with increasing $k$. Our earlier study showed that with an increasing $k$ the structure of the system remains unchanged \cite{ujjwal_mf}. Thus the contribution of the regular neighbours to the entropy does not change with $k$. However, with an increase in the number of pseudo neighbours and thus $U^{pseudo}_{k}$, the total excess entropy decreases. Thus the decrease in excess entropy obtained via the TI method can be attributed to the increase in the pseudo neighbour interactions.

\subsubsection{Vibrational entropy}
\begin{figure}[!bth]
\begin{subfigure}[h]{0.4\textwidth}
\includegraphics[width=0.9\textwidth,trim = {0 0cm 0 0.0cm},clip]{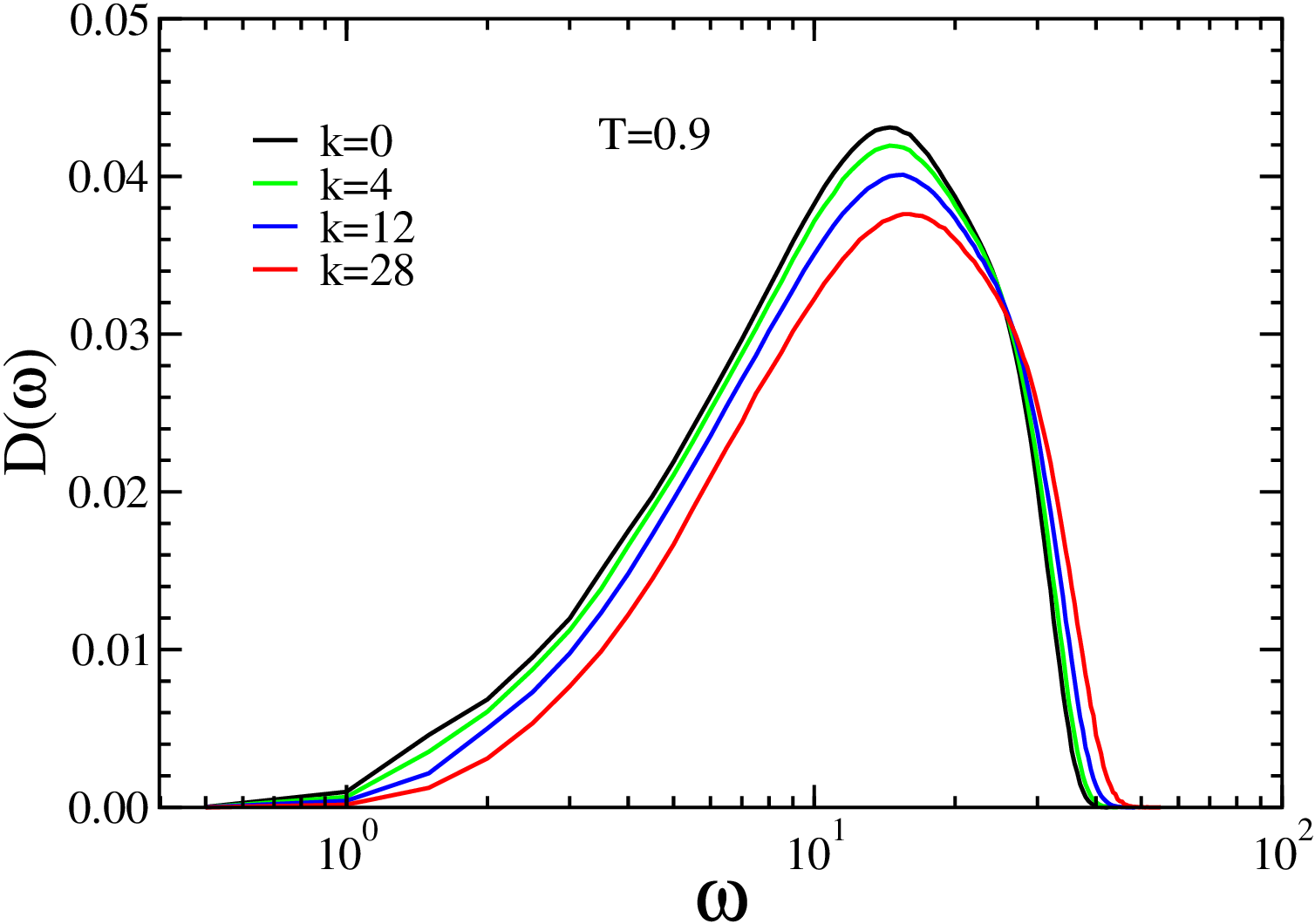}
\end{subfigure}
\caption{\emph{Vibrational density of states (VDOS), $D(\omega)$  vs. $\omega$,  for $k=0, 4, 12, 28$ systems.  With the increase in $k$, the low-frequency modes are suppressed and the whole spectrum shifts to higher frequencies.}}
\label{dos_k_0_28}
\end{figure}

We next calculate the vibrational density of states (VDOS) for different $k$ values. We find that with an increase in pseudo neighbours, there is a suppression of the low-frequency modes, and the whole spectrum moves to a higher frequency range, as shown in Fig.\ref{dos_k_0_28}. A similar effect was also seen in the high-temperature dynamics where it was shown that with the increase in the pseudo neighbours, the cage becomes stiffer and the dynamics inside the cage becomes faster \cite{ujjwal_mf}. 

\begin{figure}[!bth]
\begin{subfigure}[h]{0.4\textwidth}
\includegraphics[width=0.9\textwidth,trim = {0 0cm 0 0.0cm},clip]{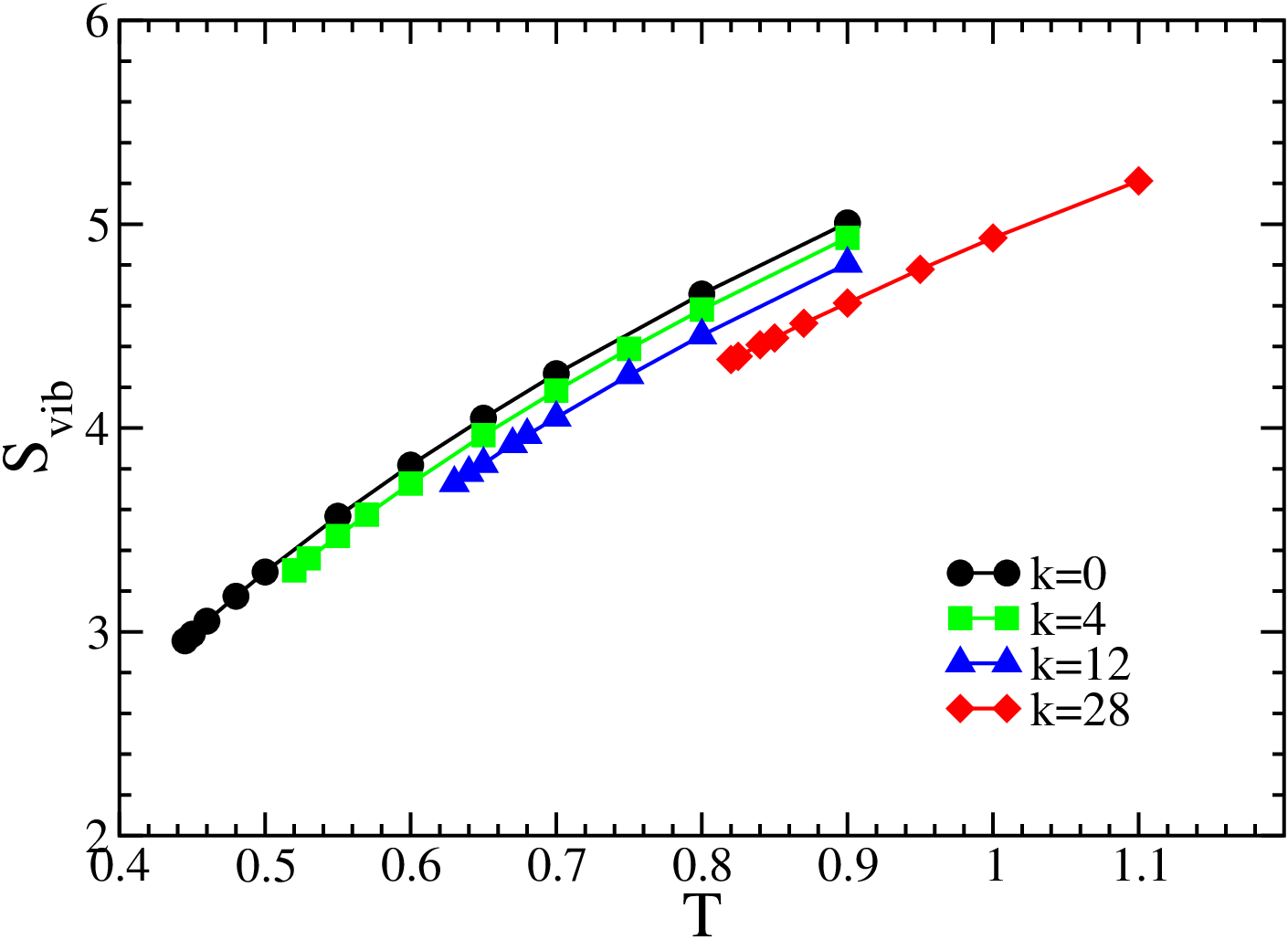}
\end{subfigure}
\caption{\emph{The vibrational entropy $S_{vib} \, vs. \, T$ for $k=0, 4, 12$ and 28 systems. With an increase in $k$, the DOS shifts to higher frequencies leading to a decrease in the vibrational entropy.}}
\label{s_vib}
\end{figure}

The temperature dependence of the vibrational entropy $S_{vib}$ (obtained from the VDOS) is plotted in Fig.\ref{s_vib}. We find that with increasing $k$, as the vibrational spectrum shifts to higher frequencies, the vibrational entropy decreases. 

\subsubsection{Configurational entropy}
\begin{figure}[!bth]
\begin{subfigure}[h]{0.4\textwidth}
\includegraphics[width=0.9\textwidth,trim = {0 0cm 0 0.0cm},clip]{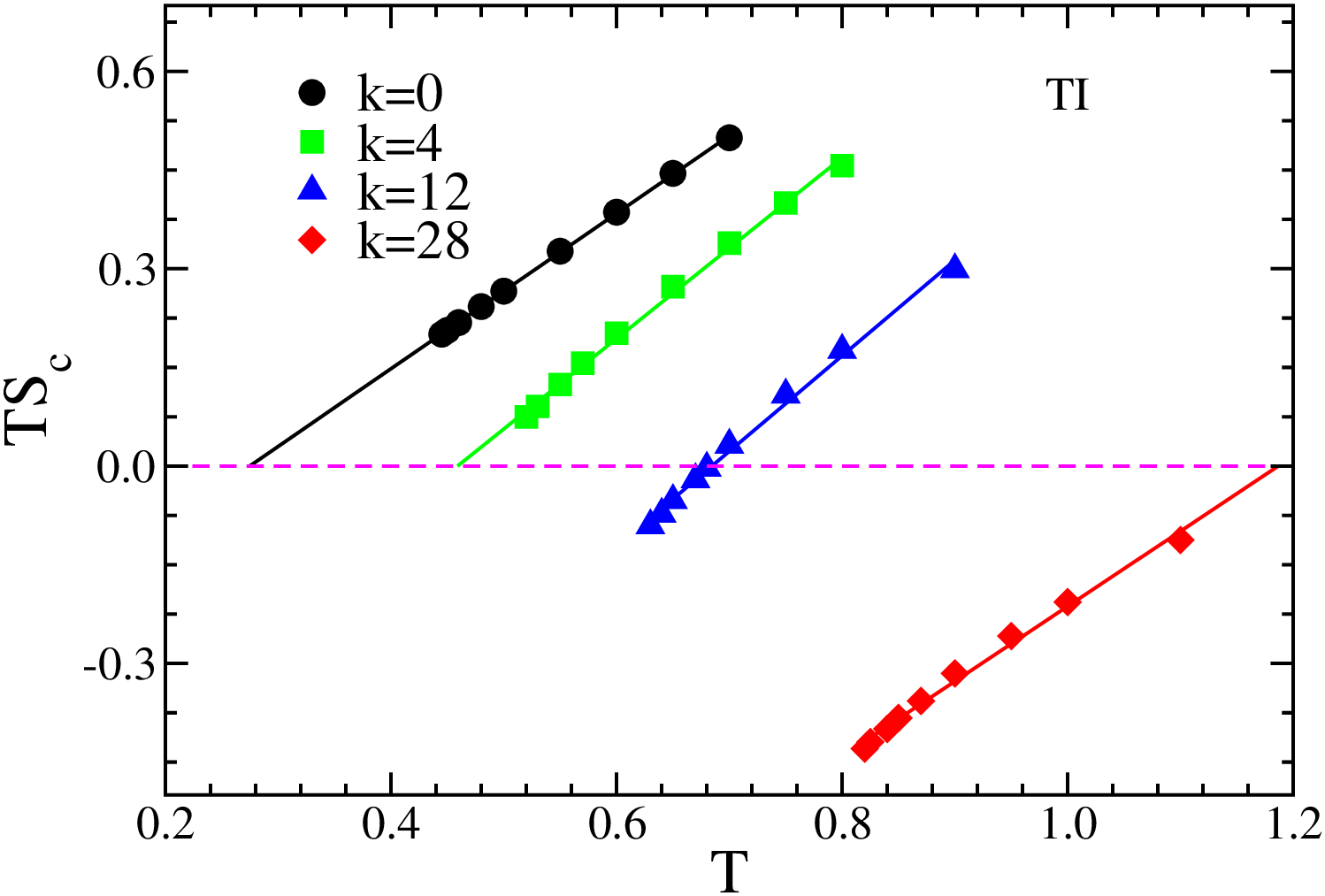}
\end{subfigure}
\caption{\emph{$TS_{c} \, vs. \, T$ for $k=0,4,12$ and 28 systems where the $S_{c}$ is calculated using the TI method. The value of the Kauzmann temperature $T_{K}^{TI}$ increases with increasing $k$. The value of $T_{K}^{TI}$ (see Table.\ref{table_compare_temp}) for the $k=28$ system is close to its onset temperature. For $k=12, 28$ systems, $T_{K}^{TI}$ values are high enough such that temperatures below $T_{K}^{TI}$ are accessible in simulation. $S_{c}$ becomes negative for such temperatures.}}
\label{TSc_vs_T}
\end{figure}

Next, we study the configurational entropy of the system. For all the systems the data is plotted below their respective onset temperatures (see Table.\ref{table_compare_temp})\cite{ujjwal_mf}. The systems follow the expected linear relationship between $TS_{c}$ and $T$ (Fig.\ref{TSc_vs_T}). The Kauzmann temperature $T^{TI}_{K}$ is obtained by fitting to $TS_{c}=K_{T}(\frac{T}{T_{K}}-1)$. We find that $T^{TI}_{K}$ increases with $k$. This is expected as in the earlier study it was found that with an increase in pseudo neighbours, the $\alpha$ relaxation time of the system appears to diverge at a higher temperature\cite{ujjwal_mf}.  However, the unphysical part of the result is the vanishing of the configurational entropy for larger $k$ systems (k=12 and 28) at comparatively high temperatures where the system can be equilibrated in simulations. Especially for the $k=28$ system, the temperature where the configurational entropy vanishes is close to the onset temperature of glassy dynamics\cite{ujjwal_mf}. The $T^{TI}_{K}$ values are listed in the Table.\ref{table_compare_temp}. In the same Table, we also list the respective $T_{0}$ values. 
For many systems, it is found that $T_{K} \simeq T_{0}$ which suggests that the slowing down of the dynamics is driven by thermodynamics \cite{Adam-Gibbs}. On the contrary, in Table.\ref{table_compare_temp} we find that the difference between the $T^{TI}_{K}$ and $T_{0}$ increases with increase in $k$.

\begin{figure}[!bth]
\begin{subfigure}[h]{0.4\textwidth}
\includegraphics[width=0.9\textwidth,trim = {0 0cm 0 0.0cm},clip]{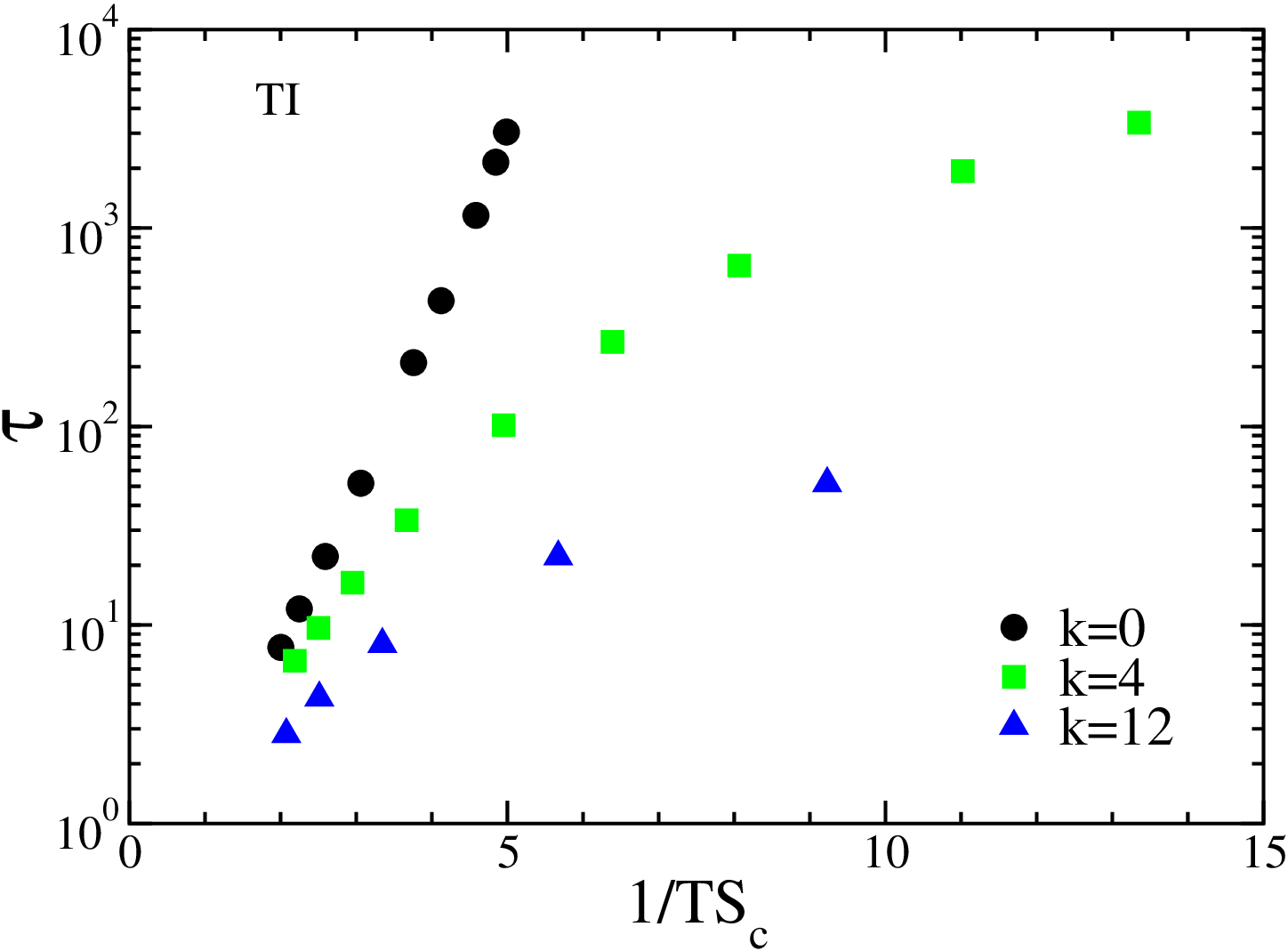}
\end{subfigure}
\caption{\emph{Testing the Adam-Gibbs relation between the relaxation time $\tau$ and $1/TS_{c}$,  for the $k=0,4$ and 12 systems.  The AG relation is obeyed for the $k=0$ system, but is violated for non-zero $k$ systems. The relaxation time $\tau$ is estimated from the self-part of the overlap function.}} 
\label{ag}
\end{figure}

 The correlation between the dynamics and thermodynamics is also given by the AG relation, $\tau=\tau_{0} \exp(-\frac{A}{TS_{c}})$. Note that this expression implies that the divergence of the relaxation time is an effect of the vanishing of the configurational entropy and if we replace the expression of $TS_{c}$ in terms of $T_{K}$ then we get back the VFT expression provided we assume $T_{K}=T_{0}$. If the system follows the AG relation then the semi-log plot of $\tau$ vs $\frac{1}{TS_{c}}$ should follow a linear behaviour which it does for most systems \cite{Adam-Gibbs,adam-gibbs_hold1,Adam-Gibbs_hold2,Adam-Gibbs_hold3,adam-Gibbs_hold4,adam-gibbs_hold5,adam-gibbs_hold6,adam-gibbs_hold7,adam-gibbs_hold8,adam-gibbs_hold9} 
 In  Fig.\ref{ag}, we study the validity of the AG relationship and find that with an increase in $k$ there is a departure from linearity. We next show that for the $k=28$ system at $T=0.82$ which is much below the $T^{TI}_{K}=1.19$, both the collective overlap function and the intermediate scattering function decay with time and reach their respective long-time values ($Q^{tot}(t\rightarrow \infty ) =0.135$ and $F(q,t \rightarrow \infty)=0)$. 
 Note that because of the introduction of the pseudo neighbours at a distance ``$L_{ij}$'', the system has more than one length scale.
 Thus to make sure that the relaxation persists at length scales that are larger and smaller than the nearest neighbour distance, we plot the intermediate scattering function at wave numbers larger and smaller than $q_{max}=\frac{2\pi}{\sigma_{max}}$ where $\sigma_{max}$ is the position of the first peak in the radial distribution function. We find that the intermediate scattering functions relax to zero at all length scales.
 Note that more than the breakdown of the AG relation which has been suggested to be a possibility \cite{Berthier_AG_hold}, 
 the fact that the dynamics shows full relaxation where the configurational entropy vanishes suggests that we need to revisit the TI method of calculating the entropy. In the TI method, we need information about the ideal gas entropy, the excess entropy, and the vibrational entropy. The vibrational entropy calculation was cross-checked by calculating it from the Fourier transform of the velocity autocorrelation function which matched the data obtained from the Hessian (See Appendix I).
 
\begin{figure}[!bth]
\begin{subfigure}[h]{0.4\textwidth}
\includegraphics[width=0.9\textwidth,trim = {0 0cm 0 0.0cm},clip]{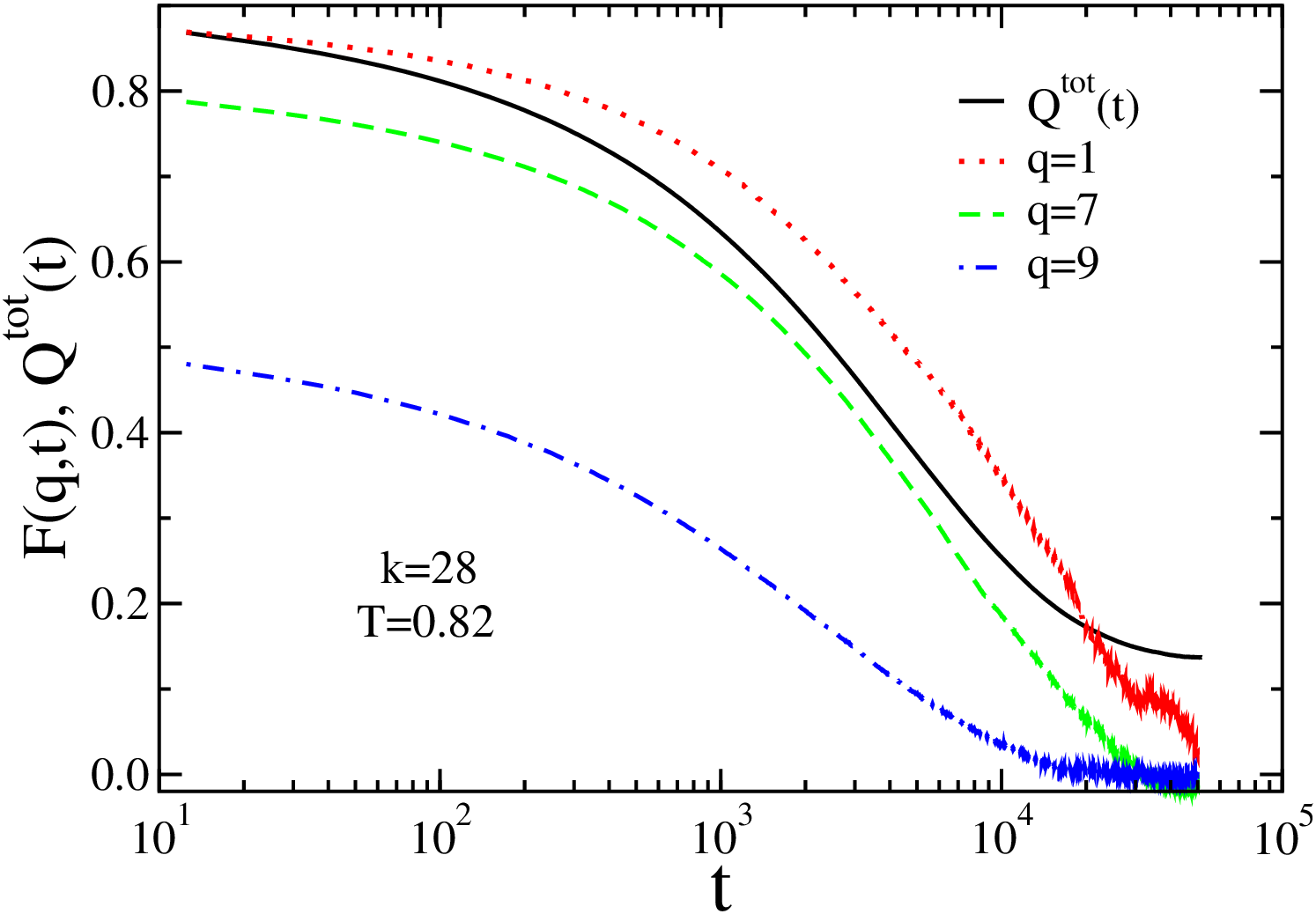}
\end{subfigure}

\caption{\emph{Time dependence of the intermediate scattering function and the collective overlap function for the $k=28$ system at a temperature $T=0.82$ which is lower than $T^{TI}_{K}$(see Table.\ref{table_compare_temp}). It shows that the self and the collective dynamics relax to their asymptotic values over time scales accessible in simulations at a temperature lower than that at which the configurational entropy vanishes.}}
\label{overlap_tot}
\end{figure}

\section{Possible reasons for the failure of the TI method}
\label{sec_discuss}

Let us first summarize the main observations made here when the entropy is calculated using TI method (i) Negative values of $S_c$ at low temperatures for large values of $k$; (ii) Full relaxation of the dynamical quantities at temperatures lower than the temperature at which $S_c$ goes to zero; (iii) Breakdown of the AG relation.
In this section, we discuss the possible failure points of the TI method. 

\subsection{Ideal gas entropy }

In the calculation of the configurational entropy  (Eq.\ref{S_c_eq}), we need the information of the ideal gas entropy. 
To make the entropy an extensive quantity we calculate the ideal gas entropy (Section \ref{ideal_gas}) by assuming the particles to be indistinguishable. However, in the mean-field system, each particle has a different set of pseudo neighbours with different $L$ values. Thus one might argue that the particles are distinguishable. 

If we assume all particles to be distinguishable {\it i e.} $m=N$, then the entropy in the thermodynamic limit will diverge (Eq.\ref{S_ideal_eq_disting}). However, for finite N, we can estimate the entropy which will increase by a factor that is proportional to $\log$(N) but independent of k. From our analysis, it appears that with an increase in 'k' the error in the entropy calculation increases. This implies that the correction  term should depend on 'k'.
Apart from the distinguishability factor, there is one other issue that can affect the ideal gas term. Here the way the interaction between a particle and its pseudo neighbour is designed restricts the particle to access a certain part of the total volume. Per pseudo neighbour this volume is a spherical region of radius $L_{ij}$. Thus in the ideal gas limit, the whole volume of the system is not accessible to a particle. The per particle inaccessible volume should increase with 'k' which will lower the entropy of the system. Thus the distinguishability factor will increase the entropy whereas inaccessible volume will decrease the entropy, the former is independent of 'k' but the latter depends on 'k'. This might appear to solve the 'k' dependence of the correction term. However, if we combine the distinguishability and inaccessible volume part then we will find that for systems with small values of 'k' the volume correction is really small and the distinguishability factor which is independent of 'k' increases the entropy by a large amount. Thus the dynamics for these systems will be similar to the k=0 system but the entropy calculated in this way will be much higher.

Another possibility is that the distinguishability is not a binary function but is a function 'k'. When we have these extra connections with the pseudo neighbours replacing particles with another one while keeping the identity of pseudo contacts the same can increase the energy of the system, and the larger the number of pseudo contacts the higher is the increase in the energy. This appears quite similar to the case of polydisperse systems with continuous polydispersity where  depending on the size range of the two particles the replacement may or may not keep the system in the same minimum \cite{ozawa_S_c_of_poly_exist}. It was argued that after particle swapping if the system remains in the same inherent structure minima then the two particles are indistinguishable and if not then they belong to different species. 
Thus to find the number of species we need to swap particle positions. Swapping particles while keeping the identity of the pseudo neighbours the same is not straightforward. The swap should make sure that in the new position of the particle none of the pseudo neighbours are within the interaction range $r_{c}$. With the increase in the number of pseudo neighbours these swaps will be mostly rejected thus making it impossible to quantify the number of species and thus the entropy.

\subsection{Excess entropy and the validity of the Rosenfeld relationship}
\begin{figure}[!bth]
\begin{subfigure}[h]{0.4\textwidth}
\includegraphics[width=0.9\textwidth,trim = {0 0cm 0 0.0cm},clip]{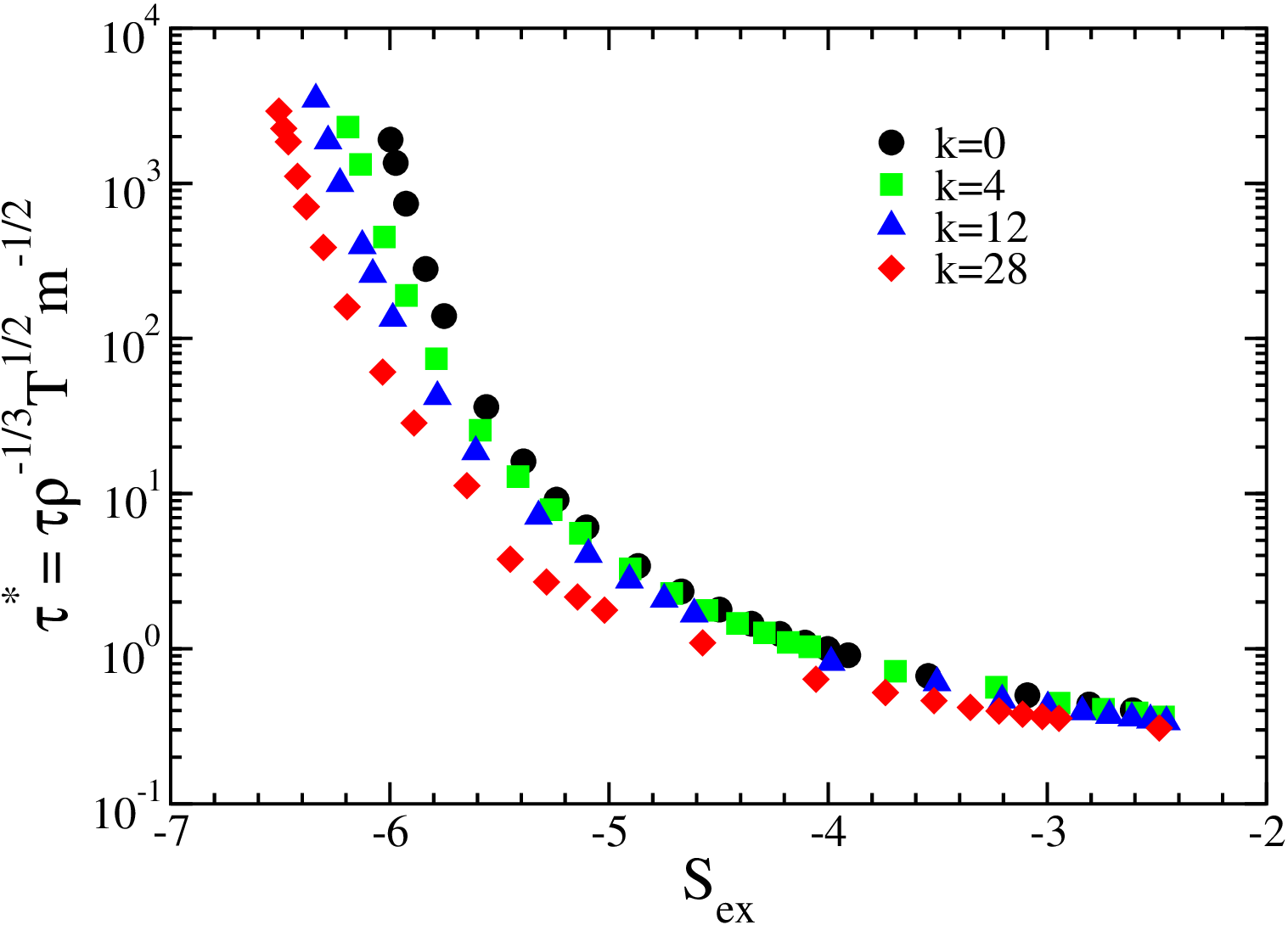}
\end{subfigure}
\caption{\emph{Scaled relaxation time vs. excess entropy. Rosenfeld scaling relation does not show universal scaling for all k systems. It deviates more from the universal scaling with increasing k.}}
\label{rosenfeld_plot}
\end{figure}

We next test the accuracy of the excess entropy value calculated via the TI method. Apart from the AG relationship which is valid in the low-temperature regime and connects the configurational entropy to the dynamics, there is another phenomenological relationship, namely the Rosenfeld relation between the excess entropy and the dynamics \cite{Rosenfeld_PRA_1977,Rosenfeld_universal_scaling}. According to the Rosenfeld relation, any dimensionless transport property will follow the excess entropy scaling. For the relaxation time it can be written as, $\tau^{*}=Rexp(-KS_{ex})$ where $\tau^{*}= \tau\rho^{-1/3}T^{1/2}m^{-1/2}$. For simple liquids, it has been found that $R \simeq 0.6$ and $K \simeq 0.8$, 
and this relationship is valid in the high-temperature regime showing a data collapse between scaled diffusion and $S_{ex}$ \cite{Charusita_JCP_structure} and also scaled relaxation time and $S_{ex}$ \cite{Atreyee_PRL,manoj_atreyee_NTW}. A recent study has also shown that scaled viscosity and diffusion coefficient for a large number of systems show a quasi universal excess entropy scaling extending over both high and low temperature regimes \cite{excess_entropy_jeppe}. In Fig.\ref{rosenfeld_plot} we plot $\tau^{*}$ vs. $S_{ex}$ for the different mean-field systems and do not find any data collapse. Thus we find a breakdown of the Rosenfeld relation and also the quasi universal excess entropy scaling \cite{excess_entropy_jeppe}.  The deviation from the Rosenfeld relationship might appear quite weak. However note that, unlike the AG relationship where we deal with the configurational entropy which has a very small value, in the Rosenfeld relationship we deal with the excess entropy which has a large value. Thus the Rosenfeld relation is not sensitive to small errors in the calculation of the entropy. 
In the mean-field system, we find that the excess entropy has a strong dependence on the number of pseudo neighbours. On the other hand, the study of the dynamics of the mean-field system showed that the interaction with the pseudo neighbours slows down the overall dynamics of the system, but has a weak effect on the structural relaxation \cite{ujjwal_mf}. Thus it appears that the role of the pseudo neighbours is not the same for the TI entropy and the dynamics. 

\subsection{Entropy using the 2PT method}
\begin{figure}[!bth]
\begin{subfigure}[h]{0.4\textwidth}
\includegraphics[width=0.9\textwidth,trim = {0 0cm 0 0.0cm},clip]{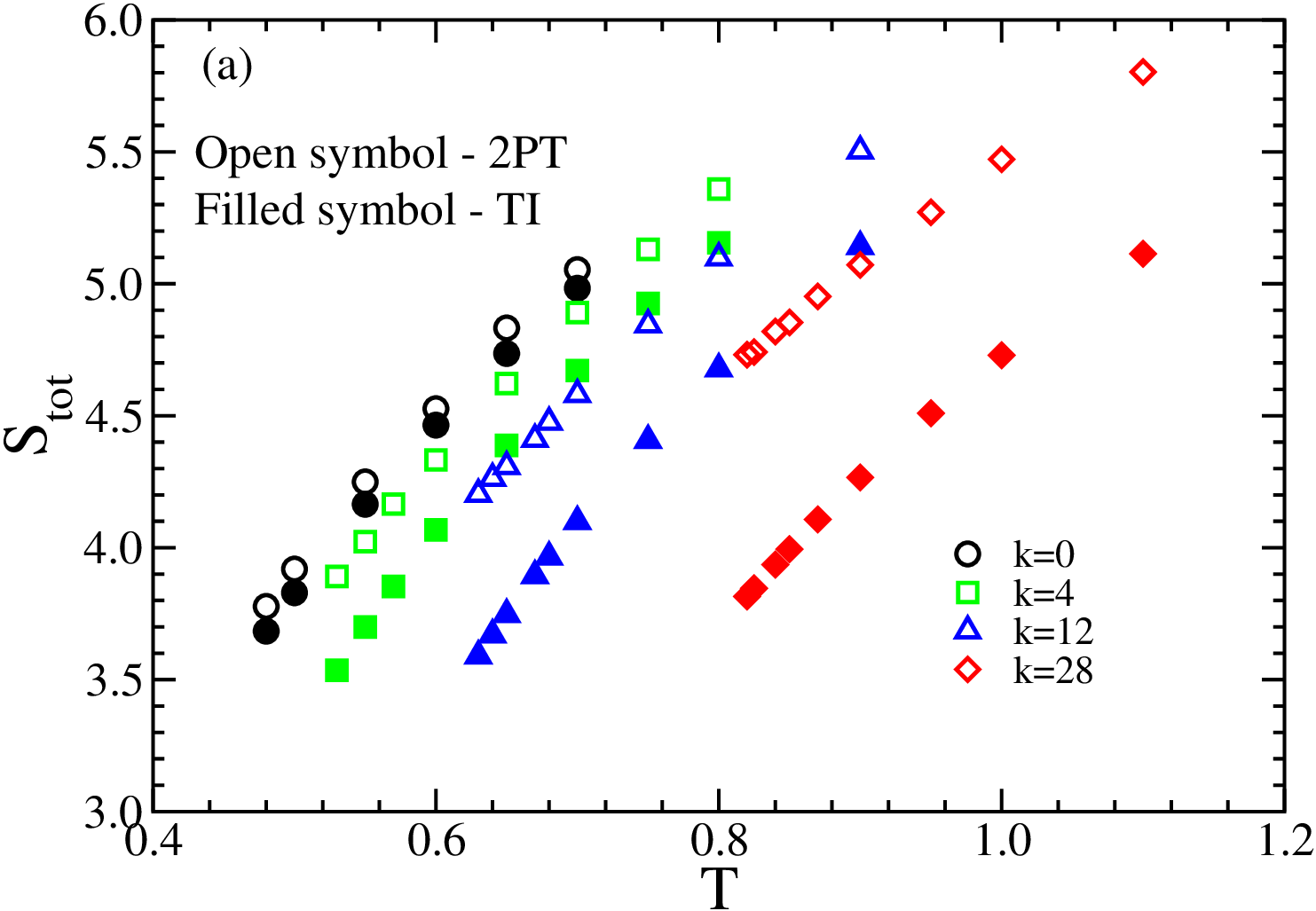}
\end{subfigure}
\begin{subfigure}[h]{0.4\textwidth}
\includegraphics[width=0.9\textwidth,trim = {0 0cm 0 0cm},clip]{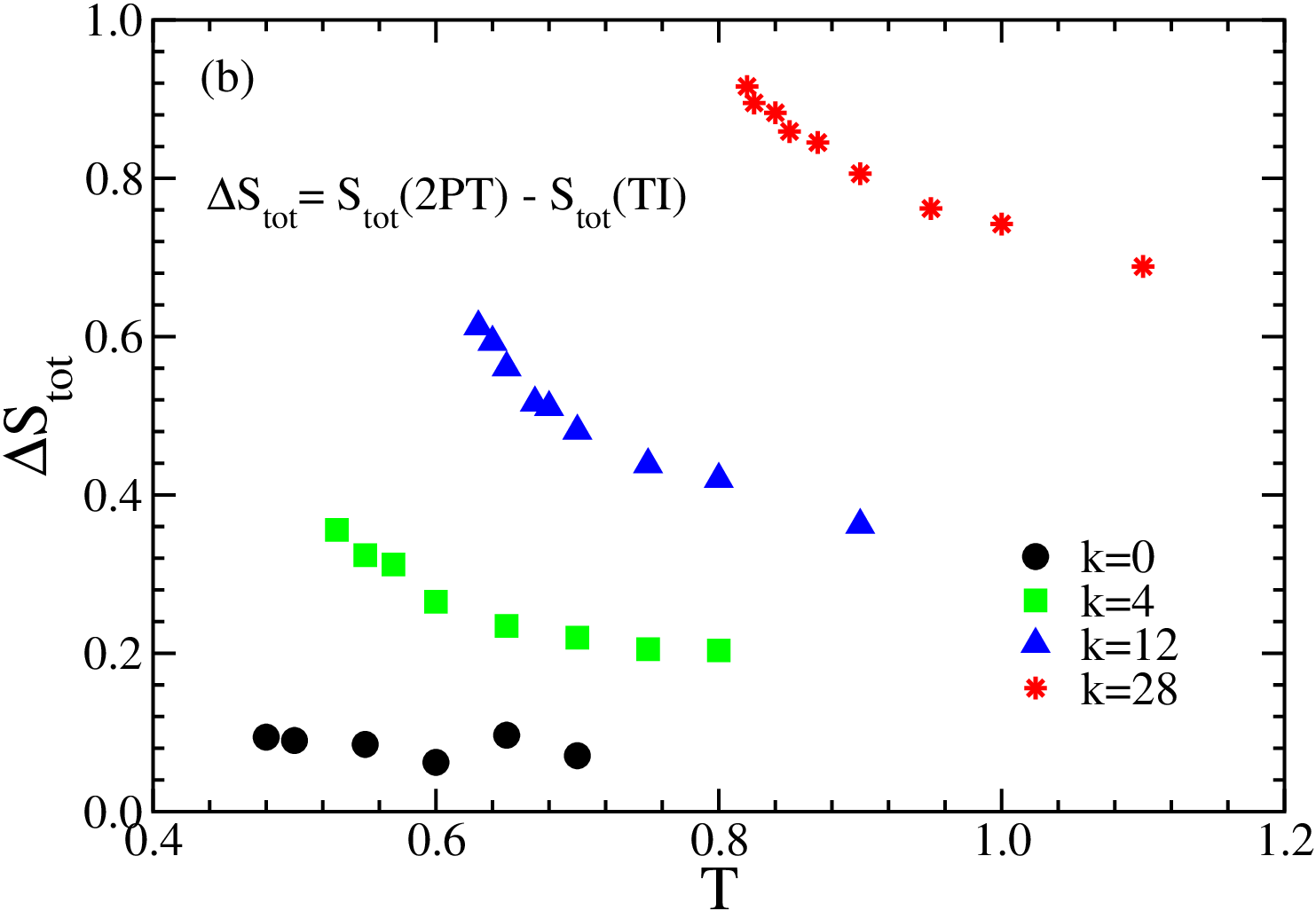}
\end{subfigure}
\begin{subfigure}[h]{0.4\textwidth}
\includegraphics[width=0.9\textwidth,trim = {0 0cm 0 0cm},clip]{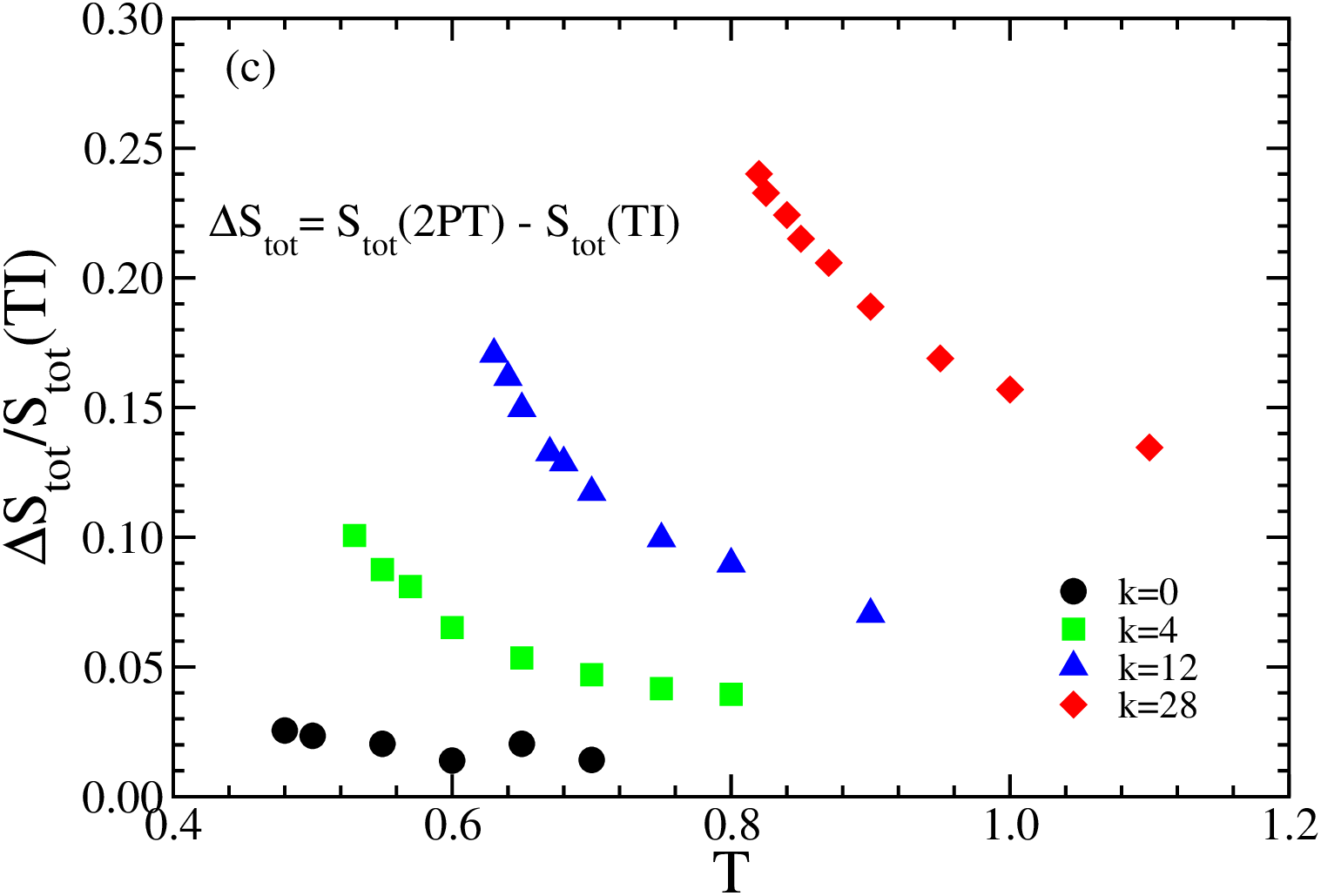}
\end{subfigure}
\caption{\emph{Comparison of the TI and 2PT methods of calculation of the entropy: (a) $S_{tot} \, vs. \, T$. Filled symbols represent results obtained from the TI method and open symbols represent those from the 2PT method. $S_{tot}$ computed by the 2PT method is higher than that by the TI method. (b) The difference in total entropy, $\Delta S_{tot}$, between 2PT and TI methods increases with increasing $k$. (c) The relative difference in the total entropy, $\frac{\Delta S_{tot}}{S_{tot}(TI)}$, between 2PT and TI methods shows similar behavior as Fig.\ref{s_tot_TI_2PT_all_k} (b).}}
\label{s_tot_TI_2PT_all_k}
\end{figure}

\begin{figure}[!bth]
\begin{subfigure}[h]{0.4\textwidth}
\includegraphics[width=0.9\textwidth,trim = {0 0cm 0 0.0cm},clip]{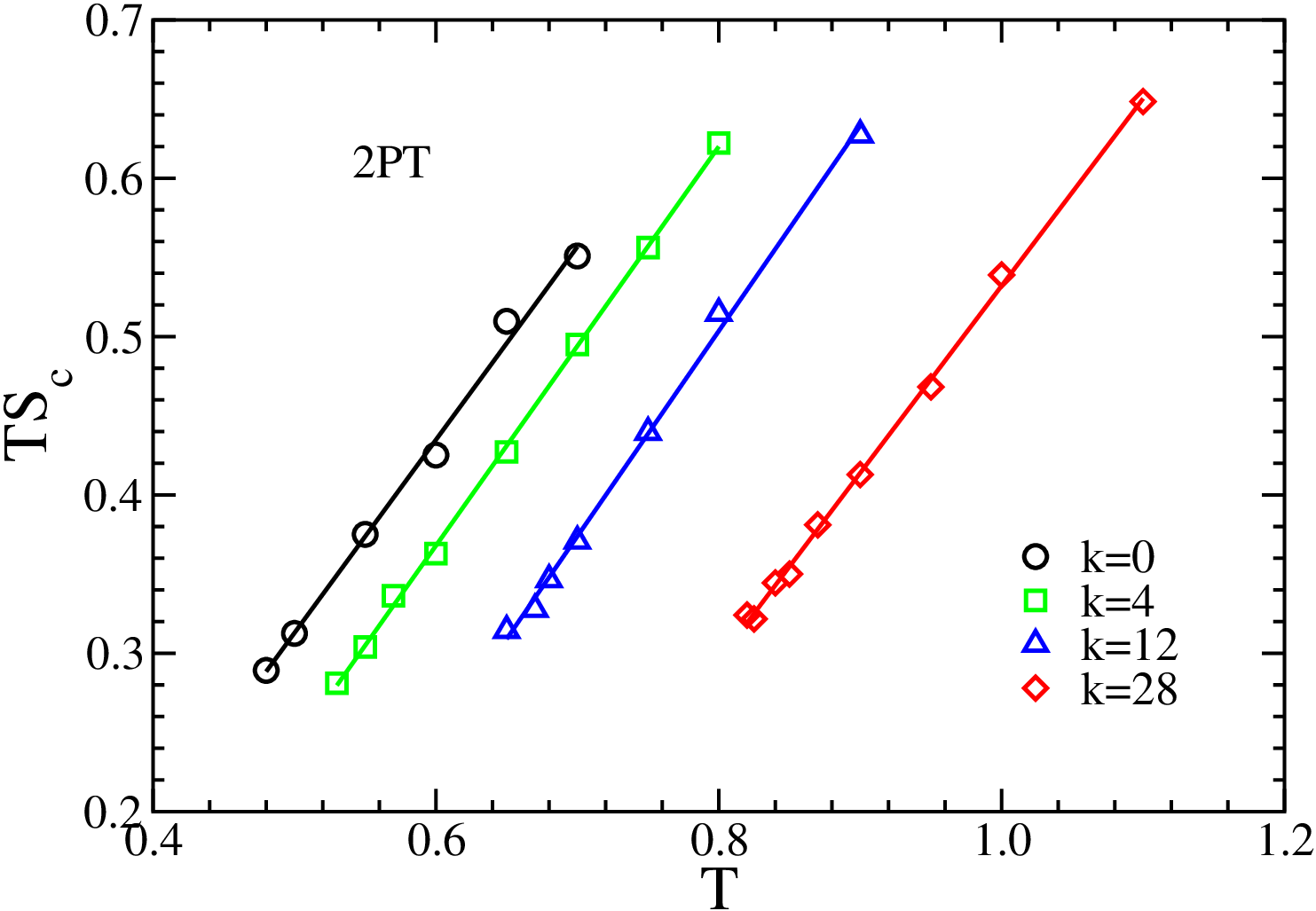}
\end{subfigure}
\caption{\emph{$TS_{c} \, vs. \, T$ for $k=0,4,12$, and 28 systems using the 2PT method. Values of $T_{K}^{2PT}$, which are close to $T_{0}$, are given in Table.\ref{table_compare_temp}.}}
\label{TSc_TI_2PT_all_k}
\end{figure}

\begin{figure}[!bth]
\begin{subfigure}[h]{0.4\textwidth}
\includegraphics[width=0.9\textwidth,trim = {0 0cm 0 0.0cm},clip]{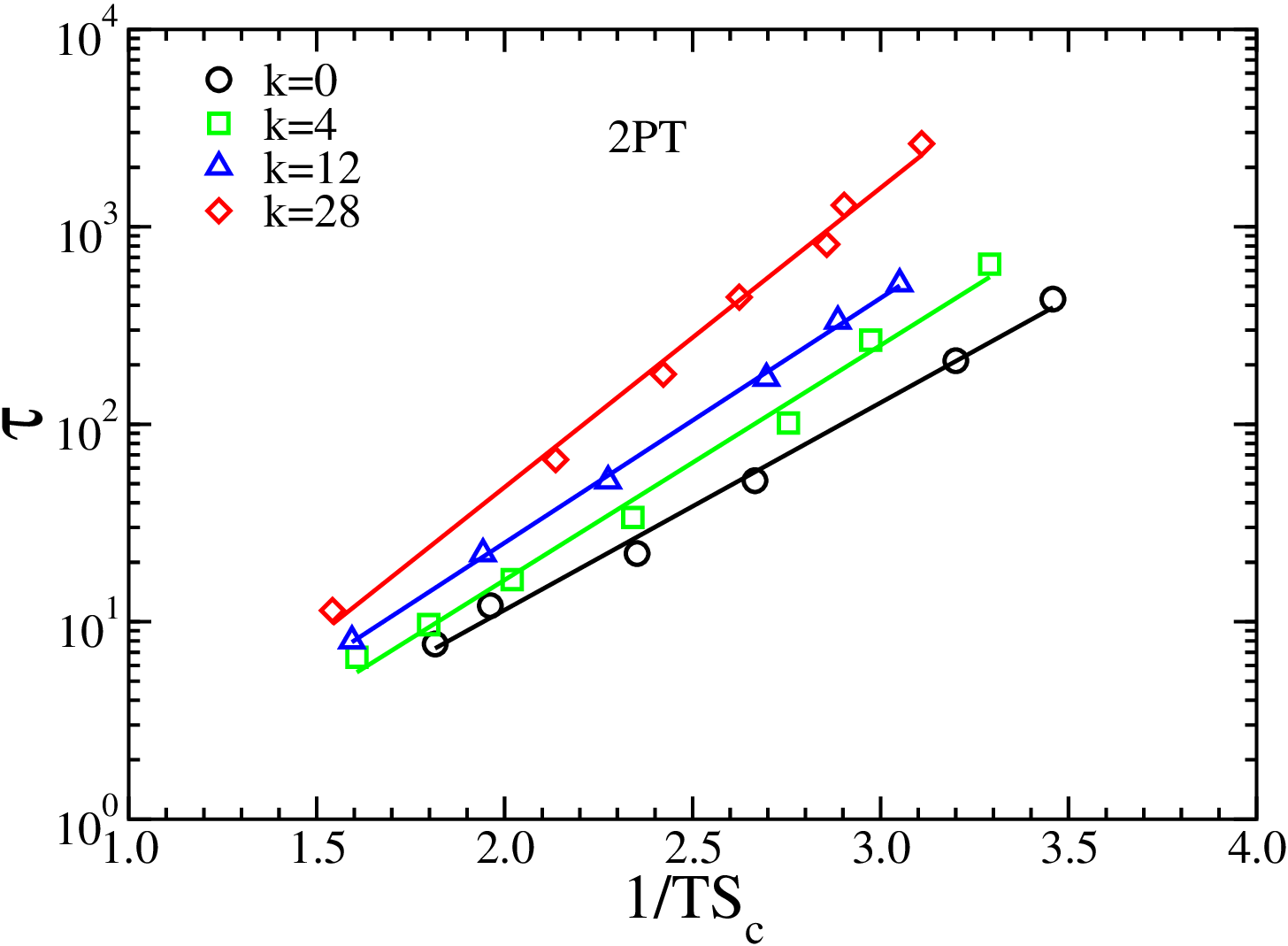}
\end{subfigure}

\caption{\emph{Testing the AG relation, $\tau$ vs. $\frac{1}{TS_{c}}$, for $k=0,4, 12$, and 28 systems with $S_c$ computed by the 2PT method. All the systems follow the AG relation in the range of temperatures studied here.} }
\label{ag_TI_2PT_all_k}
\end{figure}

Although we point out the possible sources of error in the TI method, we do not know how to correct them at present. Thus in this section, we present the results of the calculation of entropy using the 2PT method, which uses an entirely different technique. In the 2PT method, we primarily use information about the dynamics, namely the velocity autocorrelation function, to determine the entropy.
We know that the TI method works well for the regular KA model. 
Thus to validate the 2PT method, we compare it with the TI method for a regular KA system ($k=0$). As shown in Appendix I, the 2PT method works well. At temperatures close to the mode-coupling transition temperature, the 2PT method shows some deviation which is identified as arising from an averaging issue. Thus we use the results from the 2PT method in the temperature range where the upper bound is the onset temperature and the lower bound is above the respective mode-coupling theory transition temperature\cite{ujjwal_mf}. In this section, we will first compare the total entropy obtained using the 2PT method (Eq.\ref{total_2PT}) and the TI (Eq.\ref{total_TI}) method for the different mean-field systems.  As shown in Fig.\ref{s_tot_TI_2PT_all_k} the difference in total entropy between TI and 2PT method increases systematically with increasing $k$. This suggests that for this system, the TI method of calculating the entropy is not correct. 
We next study the configurational entropy as predicted by the 2PT method and its correlation with the dynamics. To calculate the configurational entropy, we need the information of the vibrational entropy, which is the same as that used in the TI method. In Fig.\ref{TSc_TI_2PT_all_k} we show the $TS_{c}$ vs T plots. We find that for all the systems $T^{2PT}_{K}$ is smaller than $T^{TI}_{K}$ and close to $T_{0}$ (see Table.\ref{table_compare_temp}).  In Fig.\ref{ag_TI_2PT_all_k} we show a semi-log plot of $\tau$ against $\frac{1}{TS_{c}}$. It clearly shows the validity of the AG relation for all the systems in the temperature range studied.


\begin{table*}
	\caption{Values of all characteristic temperatures for systems with different $k$ values.  $T_{0}$ is the VFT temperature where the $\alpha$ relaxation time diverges according to fits to the VFT equation, Eq.~\ref{eq24}. $T_{K}^{TI}$ is the Kauzmann temperature estimated from TI. $T_{K}^{2PT}$ is the Kauzmann temperature estimated from the 2PT method.}
	\begin{center}
	\addtolength{\tabcolsep}{+20.0pt}
		\begin{tabular}{ | l | l | l | l | l | l | l | l | l | l |}
			\hline
			k  & T$_{onset}$  & T$_{0}$ & T$_{K}^{TI}$ & T$_{K}^{2PT}$   \\ \hline
			0  & 0.74 $\pm 0.04$  & 0.28 & 0.28  & 0.24      \\ \hline
			4  & $0.83 \pm 0.08$  & 0.36 & 0.46  & 0.31      \\ \hline
		    12 & $1.03 \pm 0.07$   & 0.46 & 0.68  & 0.41     \\ \hline
			28 & $1.28 \pm 0.22$   & 0.61 & 1.19  & 0.55     \\ \hline

		\end{tabular}
		
		\label{table_compare_temp}
	\end{center}
\end{table*}

\section{Results for Pinned Systems}
\label{sec_pinned}

Note that in the mean-field system, the breakdown of the AG relation and also the vanishing of the configurational entropy at a temperature where the dynamics show complete relaxation is similar to what has been observed for another family of models, namely, the pinned system \cite{walter_original_pinning,smarajit_chandan_dasgupta_original_pinning,S_anh_by_walter,reply_by_chandan_dasgupata,reply_by_kob}. In the pinned system, the relaxation time obtained from single-particle dynamics remains finite at temperatures for which the configurational entropy vanishes, and there is some evidence\cite{RFOT_in_pinned_chandan_smarajit} that the relaxation time associated with the collective dynamics also remains finite at such temperatures. It has also been argued that the configurational entropy has a finite value when the vibrational entropy is calculated using an anharmonic approximation \cite{S_anh_by_walter}.  

We calculate the total entropy of the pinned system using the TI method used in earlier studies \cite{walter_original_pinning} and also given in Appendix II of the present paper. We then calculate the configurational entropy by subtracting the vibration entropy from the total entropy by taking into consideration the anharmonic contribution. As discussed in Appendix II and shown in Figs. \ref{T_Sc_anh_diff_c_fig}, \ref{AG_anh_diff_c_fig}  and Table.\ref{table_compare_temp_pin}, even after taking into consideration the anharmonic term the Kauzmann temperature $T_{K}$ appears to be high and the AG relationship is violated. 

\begin{table}
	\caption{The values of all characteristic temperatures for pinned systems with different pin concentration $c$. $T_{K}^{TI}$ is the Kauzmann temperature estimated from TI. $T_{K}^{2PT}$ is the Kauzmann temperature estimated from the 2PT method.}
	\begin{center}
	\addtolength{\tabcolsep}{+20.0pt}
		\begin{tabular}{ | l | l | l |  }
			\hline
			c  & T$_{K}^{TI}$ & T$_{K}^{2PT}$    \\ \hline
			0.00  & 0.28  & 0.24      \\ \hline
			0.05  & 0.31  & 0.30       \\ \hline
		    0.10  & 0.41  & 0.32       \\ \hline
			0.15  & 0.57  & 0.41       \\ \hline

		\end{tabular}
		\label{table_compare_temp_pin}
	\end{center}
\end{table}

\begin{figure}[!bth]
\begin{subfigure}[h]{0.4\textwidth}
\includegraphics[width=0.9\textwidth,trim = {0 0cm 0 0.0cm},clip]{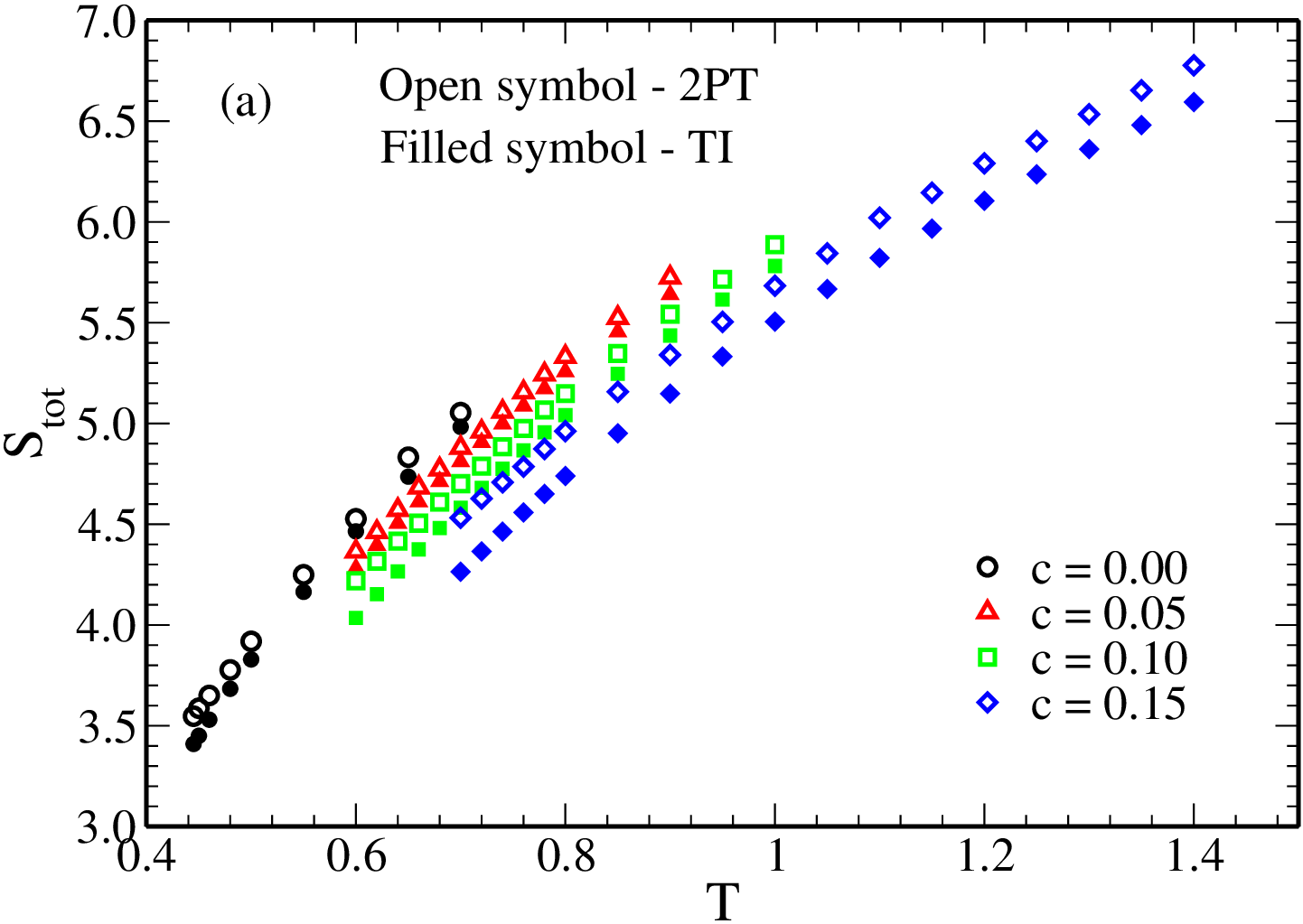}
\end{subfigure}
\begin{subfigure}[h]{0.4\textwidth}
\includegraphics[width=0.9\textwidth,trim = {0 0cm 0 0.0cm},clip]{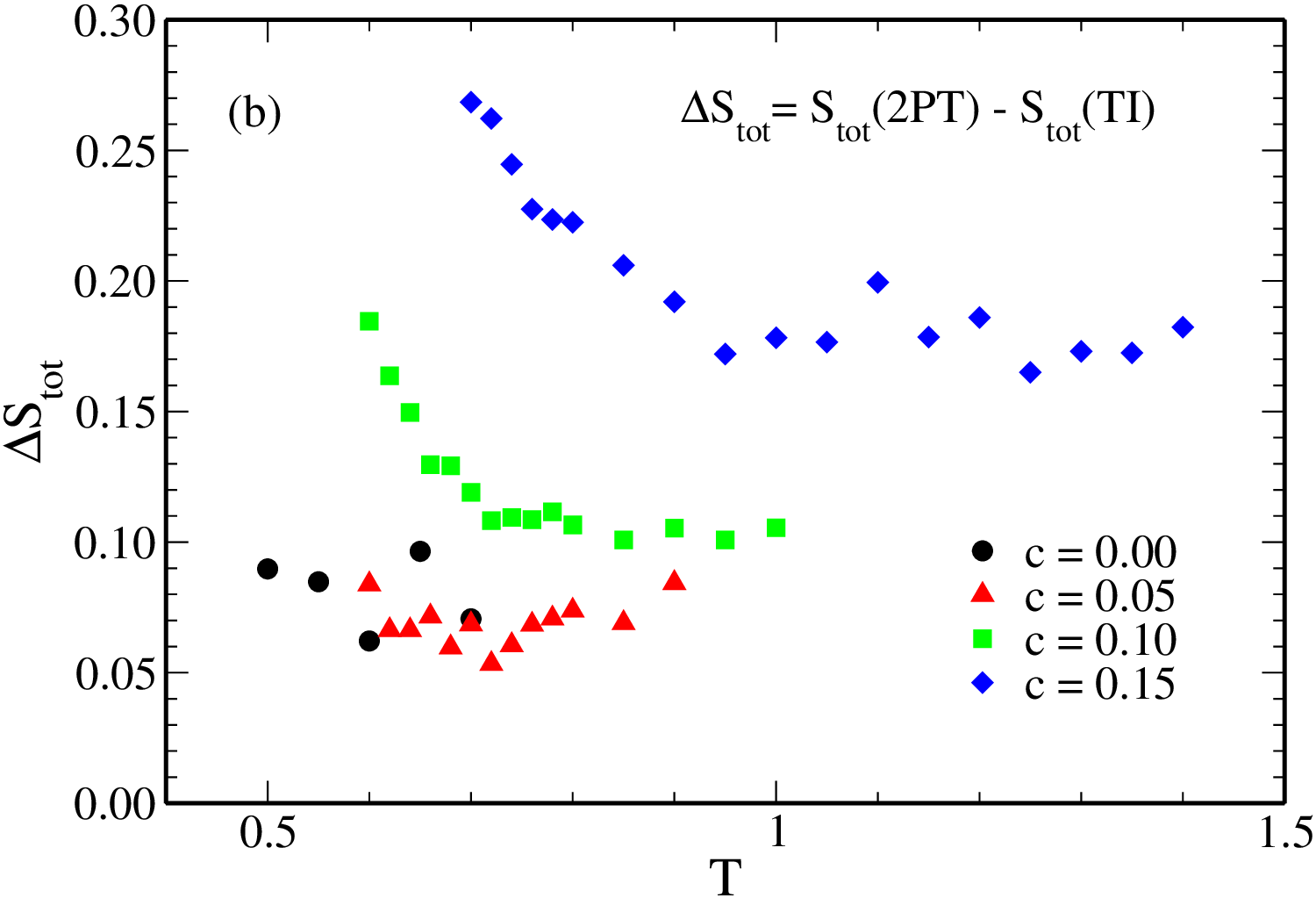}
\end{subfigure}
\begin{subfigure}[h]{0.4\textwidth}
\includegraphics[width=0.9\textwidth,trim = {0 0cm 0 0.0cm},clip]{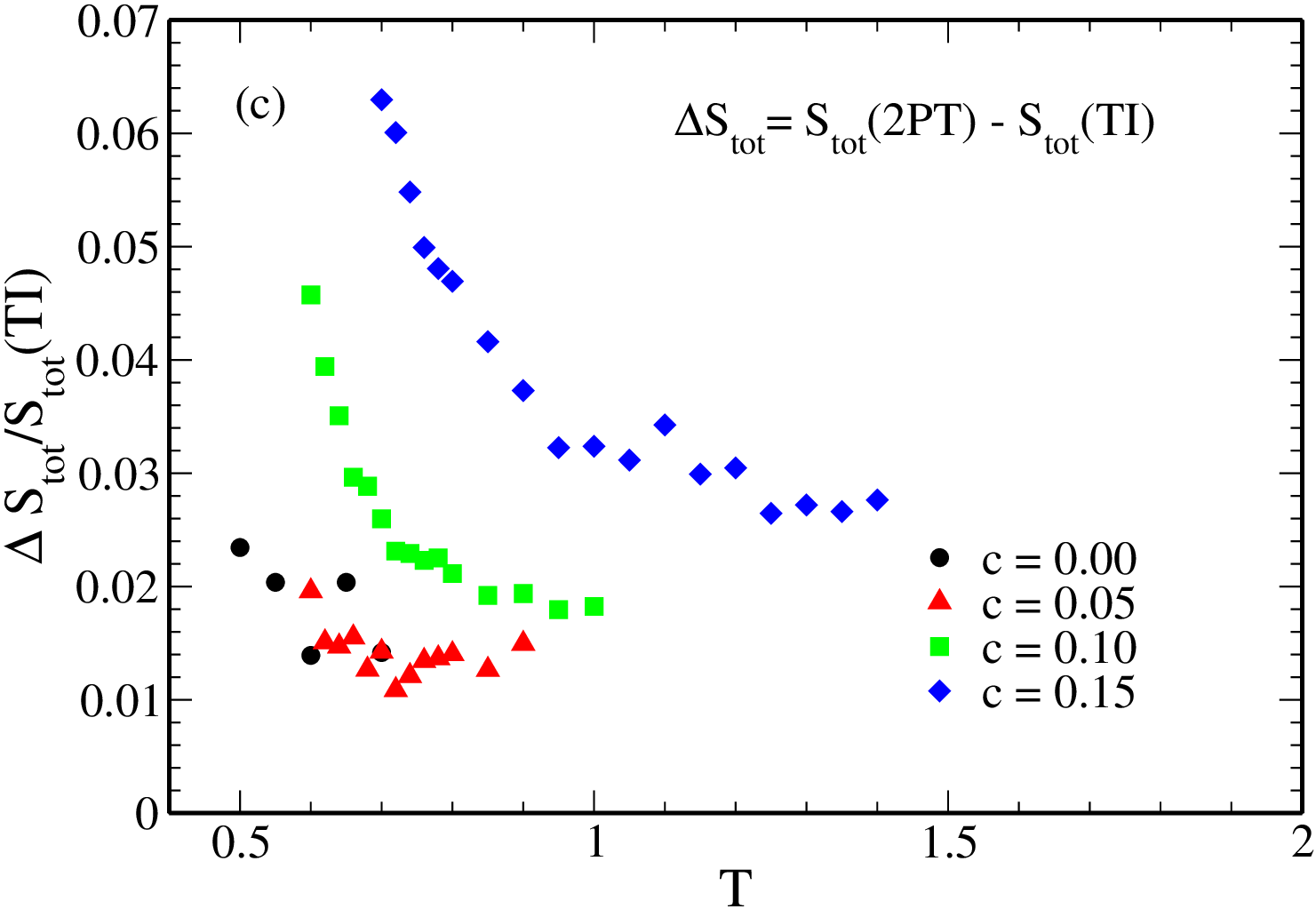}
\end{subfigure}
\caption{\emph{Comparison of the TI and 2PT methods of calculation of entropy. (a) The total entropy $S_{tot} \, vs \, T$. Filled symbols represent the results of the TI method and open symbols represent the those of the 2PT method. (b)The difference in $S_{tot}$ between 2PT and TI methods increases with increasing pinning concentration $c$. (c) The relative difference in the total entropy, $\frac{\Delta S_{tot}}{S_{tot}(TI)}$, between 2PT and TI shows similar behavior as Fig.\ref{delta_S_total_diff_c_fig} (b).}} 
\label{delta_S_total_diff_c_fig} 
\end{figure}

\begin{figure}[!bth]
\begin{subfigure}[h]{0.4\textwidth}
\includegraphics[width=0.9\textwidth,trim = {0 0cm 0 0.0cm},clip]{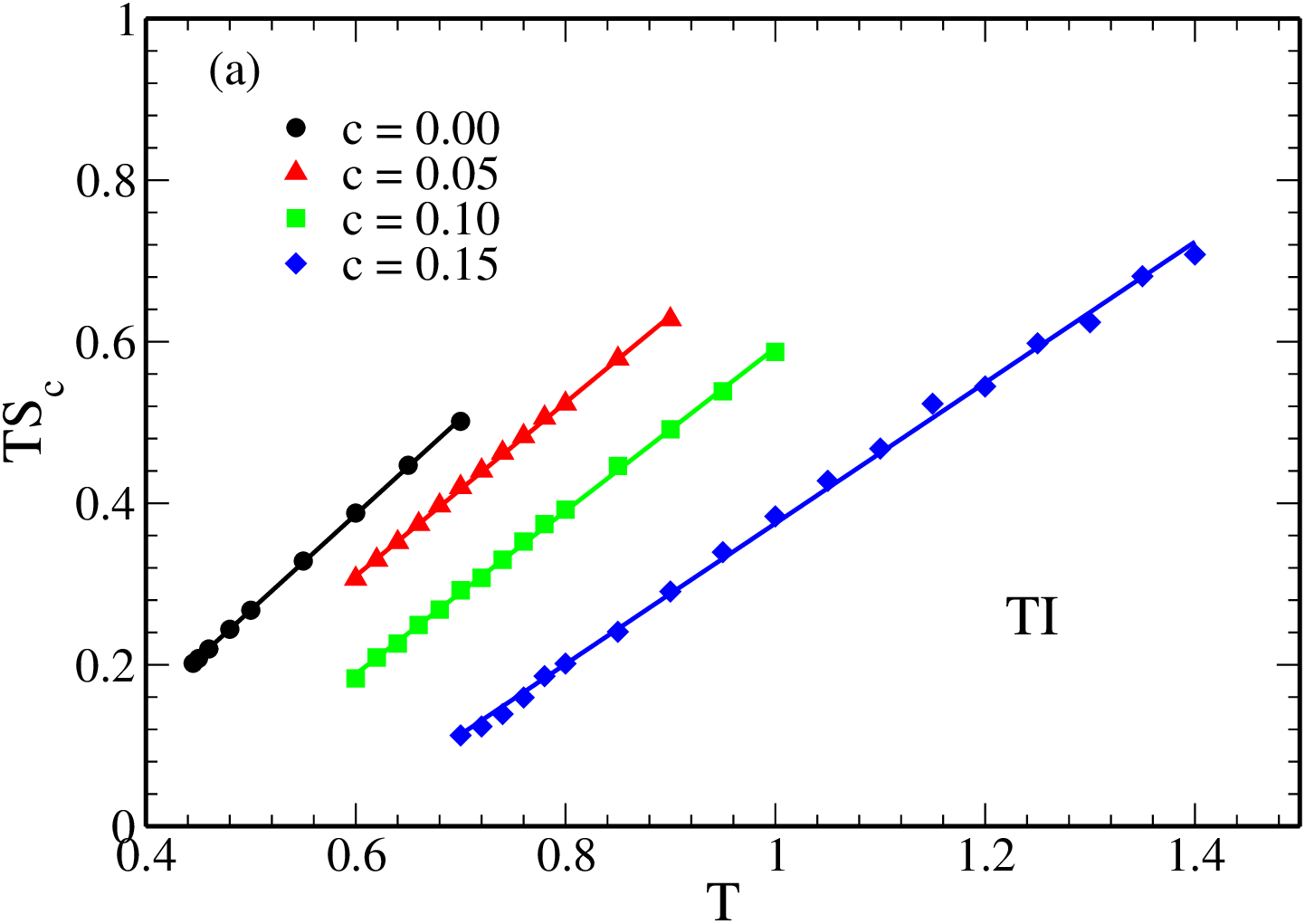}
\end{subfigure}
\begin{subfigure}[h]{0.4\textwidth}
\includegraphics[width=0.9\textwidth,trim = {0 0cm 0 0.0cm},clip]{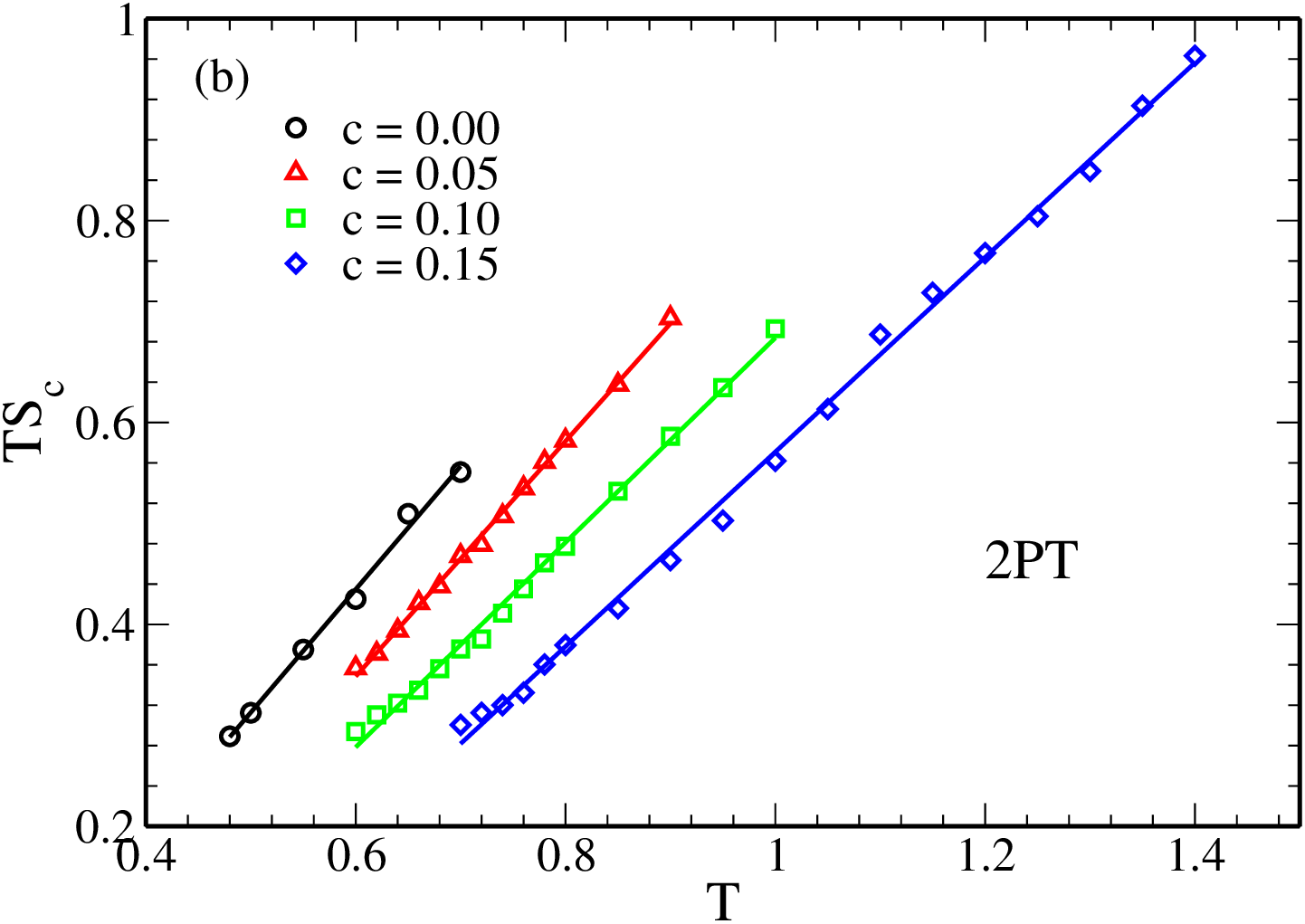}
\end{subfigure}
\caption{\emph{$TS_c \, vs. \, T$ for systems with different pinning concentrations $c=0,0.5,0.10,0.15$ using (a) the TI method and (b) the 2PT method. Both the $T^{TI}_{K}$ and the $T^{2PT}_{K}$ increase with increasing pinning concentration but $T^{2PT}_{K}<T^{TI}_{K}$, see Table.\ref{table_compare_temp_pin}.}}
\label{T_Sc_diff_c_fig}
\end{figure}

Given the success of the 2PT method in determining the entropy for the mean-field system, we apply it for the pinned system and compare it with the TI method. In Fig.\ref{delta_S_total_diff_c_fig} we plot (a) the total entropy obtained using two different methods and in Fig.\ref{delta_S_total_diff_c_fig} (b) their differences, for three different pinning densities and (c) the relative difference. For comparison, we also show the KA system with no pinning which is the same as the KA system with $k=0$. Similar to that observed in the mean-field system we find a difference between the entropy calculated via the 2PT and the TI methods that increases systematically with pinning. We next calculate the configurational entropy as predicted by the two methods and plot the temperature dependence of $TS_c$ in Fig.\ref{T_Sc_diff_c_fig}. Both methods predict positive Kauzamnn temperatures for each system and similar to the case of mean-field systems, the Kauzmann temperature predicted by the 2PT method is lower than that by the TI method, see Table.\ref{table_compare_temp_pin}. In this calculation, we have used the harmonic approximation for the vibrational entropy. The anharmonic approximation will equally affect both the 2PT and TI entropy values and the plots are given in Appendix II.

\begin{figure}[!bth]
\begin{subfigure}[h]{0.4\textwidth}
\includegraphics[width=0.9\textwidth,trim = {0 0cm 0 0.0cm},clip]{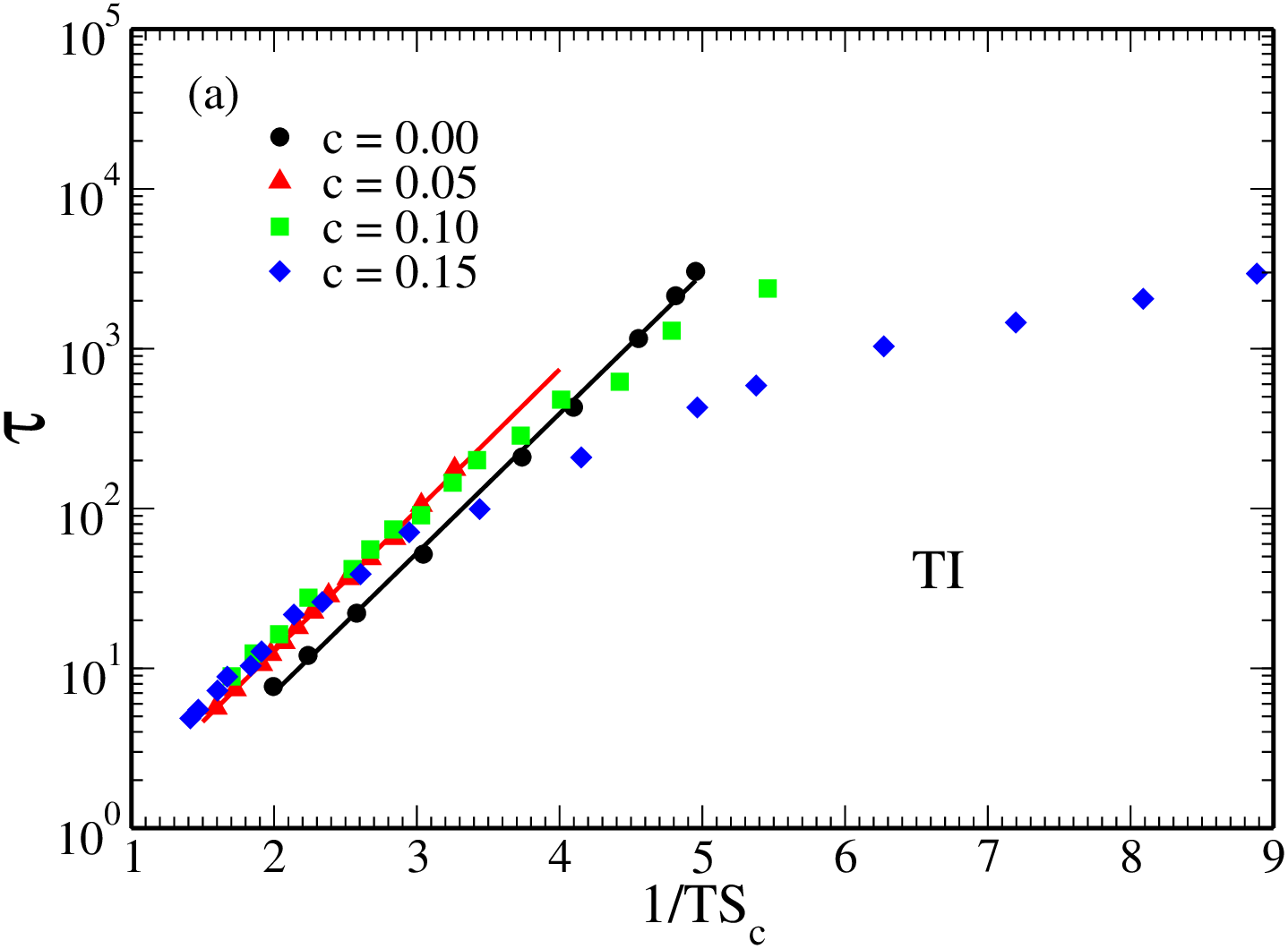}
\end{subfigure}
\begin{subfigure}[h]{0.4\textwidth}
\includegraphics[width=0.9\textwidth,trim = {0 0cm 0 0.0cm},clip]{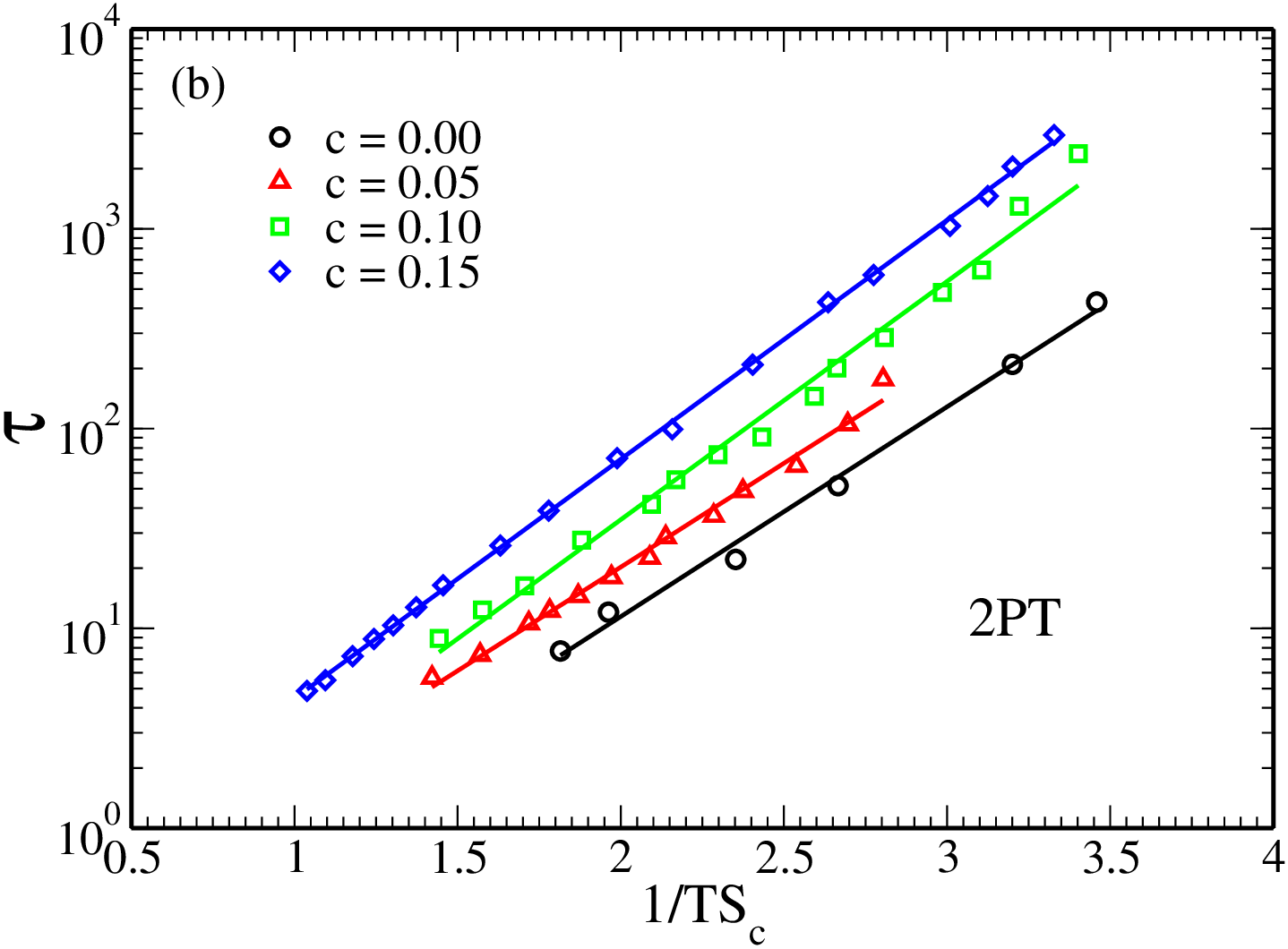}
\end{subfigure}
\caption{\emph{Testing the AG relation between $\tau$ vs. $\frac{1}{TS_{c}}$ for $c=0,0.5,0.10,0.15$ systems using (a) the TI method (b) the 2PT method. In the temperature range studied here, the AG relation is violated for $c=0.1$ and $c=0.15$ when $S_c$ is calculated using the TI method. However, the AG relation holds for all $c$ when $S_c$ is calculated via the 2PT method.}}
\label{AG_diff_c_fig}
\end{figure}

Next, we need to understand if the lowering of the $T_{K}$ value in the 2PT method is sufficient to describe the dynamics via the AG relationship. In Fig.\ref{AG_diff_c_fig} we show semi-log plots of $\tau$ vs. $\frac{1}{TS_{c}}$ where the entropy is calculated using the 2PT and the TI methods. The TI method shows a strong breakdown of the AG relation for $c=0.1$ and $c=0.15$, whereas the 2PT method clearly follows the AG relation for all $c$.

Unlike the mean-field system where the source of error in the TI method can come from the ideal gas calculation, in the pinned system there is no such possibility.  
In a recent study, it was found that although for unpinned systems the local dynamics correlates with the local pair excess entropy, with an increase in the pinning density such correlation disappears \cite{paddy}. Note that the pair excess entropy contributes to about 80 $\%$ of the total excess entropy. Thus this result is similar in spirit to the breakdown of the Rosenfeld relationship found here for the mean-field system.

\section{Conclusion}
\label{sec_conclusion}
Recently some of us have developed a model for glass-forming liquid whereby changing a parameter the system can continuously switch from a standard three-dimensional liquid to a fully connected mean-field like system \cite{ujjwal_mf}. The parameter is $k$, the number of additional particle-particle interactions that are introduced per particle on top of the regular interactions in the system. With increasing $k$, the structure and the dynamics were studied which showed more mean-field like behaviour at higher $k$ values. The present work aims to study the thermodynamics of the system and understand its correlation with the dynamics. 
To study thermodynamics, we first calculate the entropy using the well-known TI method \cite{sastry-onset}. We then study the correlation of the entropy with the dynamics. 
This model shows super-Arrhenius dynamics similar to conventional glassy liquids \cite{ujjwal_mf}, suggesting that the RFOT description should apply. However, we find that the relaxation times calculated from both single-particle and collective dynamics remain finite at temperatures where the configurational entropy vanishes. This is different from the prediction of RFOT and the behavior seen in conventional glass-forming liquids for which the (extrapolated) values of $T_K$ and $T_0$ are found to be close to each other \cite{Adam-Gibbs,Berthier_AG_hold,Berthier_S_c}. We discuss the possible source of error in the TI method of calculation of the entropy for the mean-field system. However, at this point, we do not know how to modify the TI method to correctly calculate the entropy of these model systems.

We thus use another technique namely the 2PT method to calculate the entropy. The 2PT method assumes that a liquid can be represented as partially a gas and partially a solid and this fraction is a function of the thermodynamic parameters of the system and also of the size of the particles. The 2PT method has been extensively used to calculate the entropy for many systems, mostly in the high-temperature regime \cite{lin2012two,lin2003two}. In recent work, this method was also extended to lower temperatures \cite{moid2021}. We find that for the KA system at $k=0$, both the 2PT method and the TI method provide similar results. We then compare the total entropy calculated by the 2PT method with that by the TI method for different mean-field systems. We find that the difference between the entropy values obtained in the two methods systematically increases with increasing $k$. We also find that the entropy calculated via the 2PT method describes the dynamics quite well and confirms the RFOT prediction.

The results of the mean-field systems appear quite similar to that of the pinned particle system studied earlier \cite{walter_original_pinning}. In the pinned system the self-part of the density correlation function decays to zero at temperatures where $S_c$ obtained from the TI method goes to zero \cite{smarajit_chandan_dasgupta_original_pinning}.  Given the success of the 2PT method in calculating the entropy of the mean-field system, we apply it to calculate the entropy of the pinned system. Interestingly we find that similar to the mean-field system the difference between the entropy calculated via 2PT and TI methods systematically increases with pinning. The entropy obtained via the 2PT method seems to explain the temperature dependence of the relaxation time obtained from the self overlap function well and the RFOT prediction remains valid.

Thus our analysis suggests that for a certain class of systems, the TI method in its current form fails to predict the correct value of the entropy.
At this point, we are unable to comment exactly why the TI method which has a microscopic basis fails whereas the 2PT  method which is in a way heuristic in nature succeeds in predicting the dynamics. Also, the possible source of error in the TI method of entropy calculation for the two different systems may or may not be the same. 
In the mean-field system the source of error in the entropy calculation using the TI method can be the ideal gas term as the particles due to their fixed set of pseudo neighbours can appear to be distinguishable and also the total volume of the system might not be accessible to the particles even at infinite temperature limit. However, no such error in the ideal gas term is expected in the pinned system. On the other hand, the mean-field system shows a breakdown of the Rosenfeld scaling when the excess entropy is calculated using the TI method. A recent study has shown that for the pinned system the correlation between the local pair excess entropy and the dynamics breaks down \cite{paddy}. These two results appear similar in spirit. The excess entropy calculation only depends on the interaction between particles. Thus for the mean-field system, we may be overestimating the interaction between the particle and its pseudo neighbours and for the pinned system between the unpinned and the pinned particles. This conjecture needs to be tested and more such systems need to be studied to understand the role of interaction in the estimation of entropy using the TI method.


\textbf {Appendix I: Comparison of 2PT and TI method for KA model}

For a binary system in the 2PT method of entropy calculation, we need to provide the information of the partial volume fraction which can be calculated as \cite{lin2012two},
\begin{equation}
   \bar{V_{i}} = \frac{\sigma_{i}^3}{\sum_{j}x_{j}\sigma_{j}^3} \frac{V}{N}
    \label{v_a_v_b_eq}
\end{equation}
\noindent where, $V_{i} = \bar{V_{i}} N_{i}$.

Partial volume fraction depends on the radii of the particles. In the KA system, the diameter of the $A$ and $B$ particles are 1 and 0.88. However, the potential in the KA model is designed in such a way that it allows interpenetration between the $A$ and the $B$ particles ($\sigma_{AB}<(\sigma_{A}+\sigma_{B})/2$). Thus if we assume that the $B$ particles are surrounded by all $A$ particles then the effective diameter of a $B$ particle will be 0.6. To understand the role of partial volume fraction on the entropy we calculated  $S_{tot}$ from the 2PT method, assuming the $B$ particle diameter to be 0.8 and 0.6. We find that at high temperatures the 0.6 value provides a better result but at low temperatures, the entropy is almost independent of the small changes in the partial volume fraction. Thus for these systems, we assume the diameter of the $B$ particles to be 0.6.

\begin{figure}[!bth]
\begin{subfigure}[h]{0.4\textwidth}
\includegraphics[width=0.9\textwidth,trim = {0 0cm 0 0.0cm},clip]{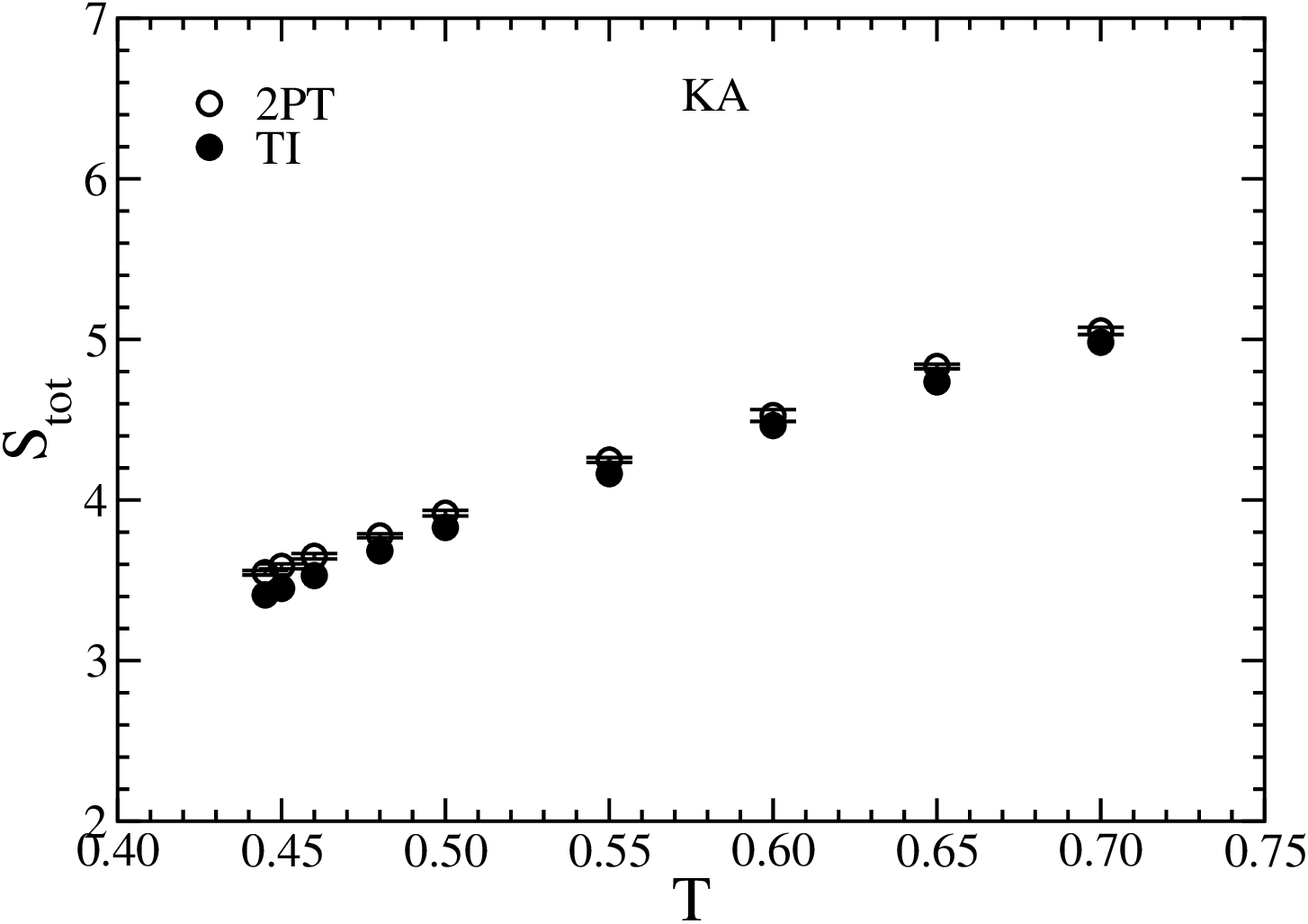}
\end{subfigure}
\caption{\emph{$S_{tot} \, vs. T$ for the KA model using the TI and the 2PT method. The two methods agree reasonably well. A small systematic deviation in the low-temperature regime is due to limited averaging possible for the 2PT method, see Fig.\ref{diff_frame_length_T_0.45}.}}
\label{s_tot_TI_2PT}
\end{figure}

\begin{figure}[!bth]
\begin{subfigure}[h]{0.4\textwidth}
\includegraphics[width=0.9\textwidth,trim = {0 0cm 0 0.0cm},clip]{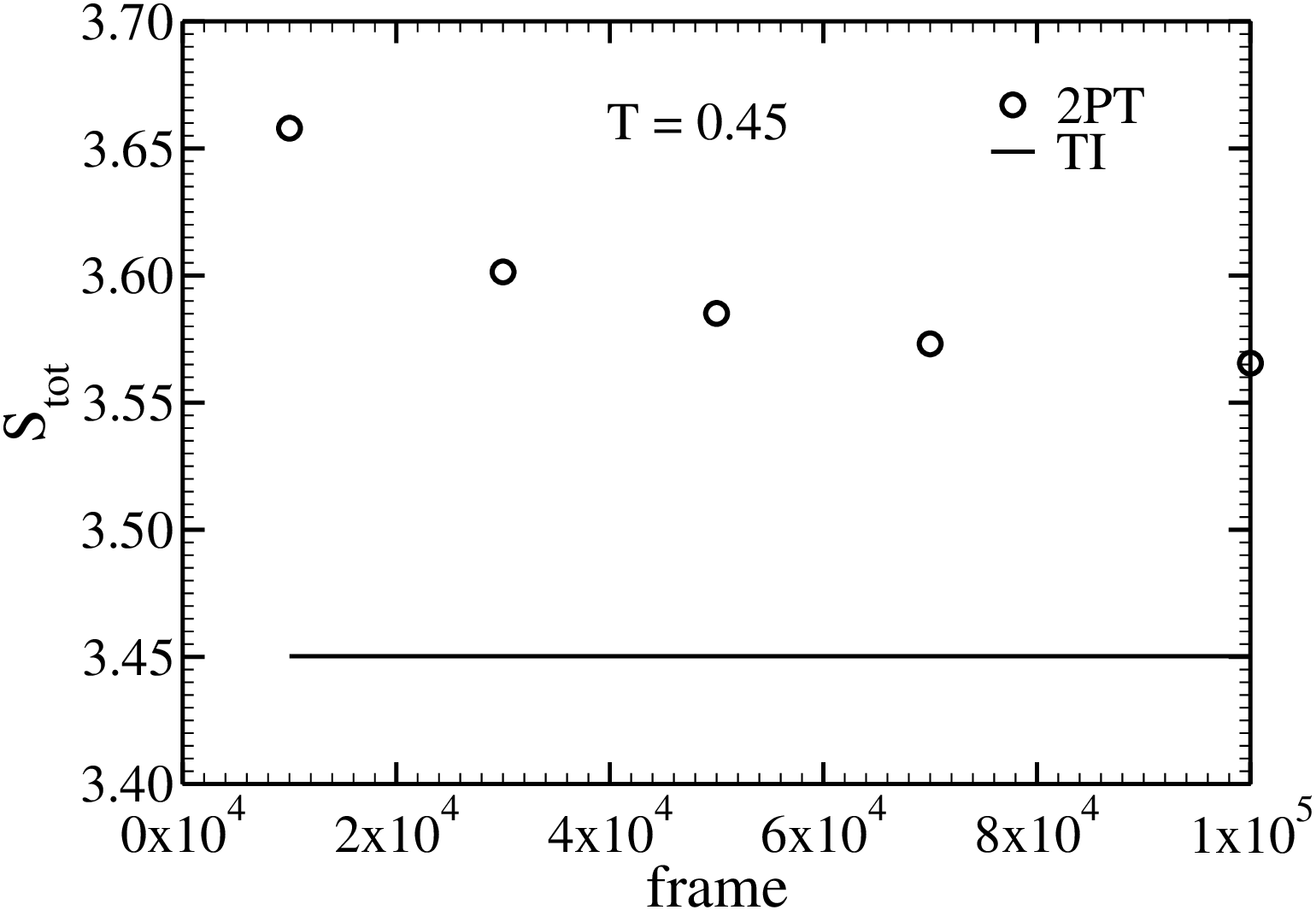}
\end{subfigure}
\caption{\emph{The total entropy via the 2PT method as a function of the number of time frames over which the velocity autocorrelation function is integrated to obtain the spectral density at a low temperature $T = 0.45$. For comparison, we also plot the entropy value obtained using the TI method. The difference decreases with increasing time interval, but the rate of convergence becomes slower at longer times.}}
\label{diff_frame_length_T_0.45}
\end{figure}

We compare the total entropy of the system as estimated from the TI \cite{sastry-nature} and from the 2PT \cite{lin2003two} methods. Fig.\ref{s_tot_TI_2PT} shows that the $S_{tot}$ obtained from TI and 2PT methods have similar values. The error bar for the 2PT data is estimated from a set of ten runs at each temperature. We find some deviation in the low temperature. At low temperatures as the dynamics become slow, we need longer runs to get a converged DOS. Fig.\ref{diff_frame_length_T_0.45} shows the effect the time step has on the value of total entropy at lower temperatures.  With an increase in time step the entropy value approaches the value calculated using the TI method. However, at longer times, the slope of the curve decreases.

\begin{figure}[!bth]
\begin{subfigure}[h]{0.4\textwidth}
\includegraphics[width=0.9\textwidth,trim = {0 0cm 0 0.0cm},clip]{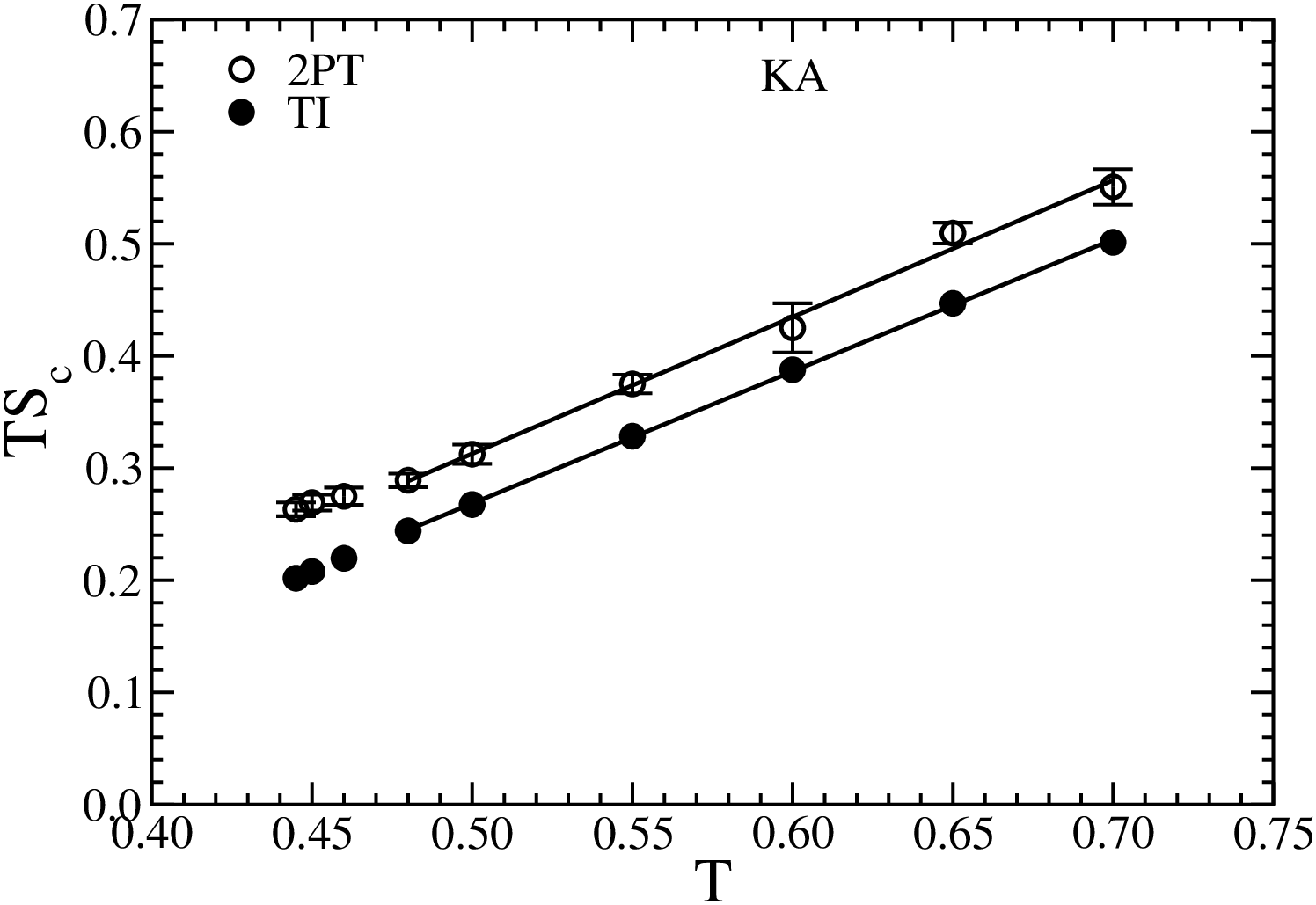}
\end{subfigure}
\caption{\emph{$TS_{c} \, vs. \, T$ for the KA model using the TI and the 2PT methods. The value of $T_{K}$ estimated by the two methods are similar ($T^{TI}_{K}$=0.27, $T^{2PT}_{K}$=0.24).}}
\label{TSc_TI_2PT}
\end{figure}

Configurational entropy, $S_{c}$ obtained in the two different methods is plotted in Fig.\ref{TSc_TI_2PT}
We find that the values of Kauzmann temperature ($T_{K}$) using two different methods are close which validates the applicability of the 2PT method for the calculation of the configurational entropy. 

\begin{figure}[!bth]
\begin{subfigure}[h]{0.4\textwidth}
\includegraphics[width=0.9\textwidth,trim = {0 0cm 0 0.0cm},clip]{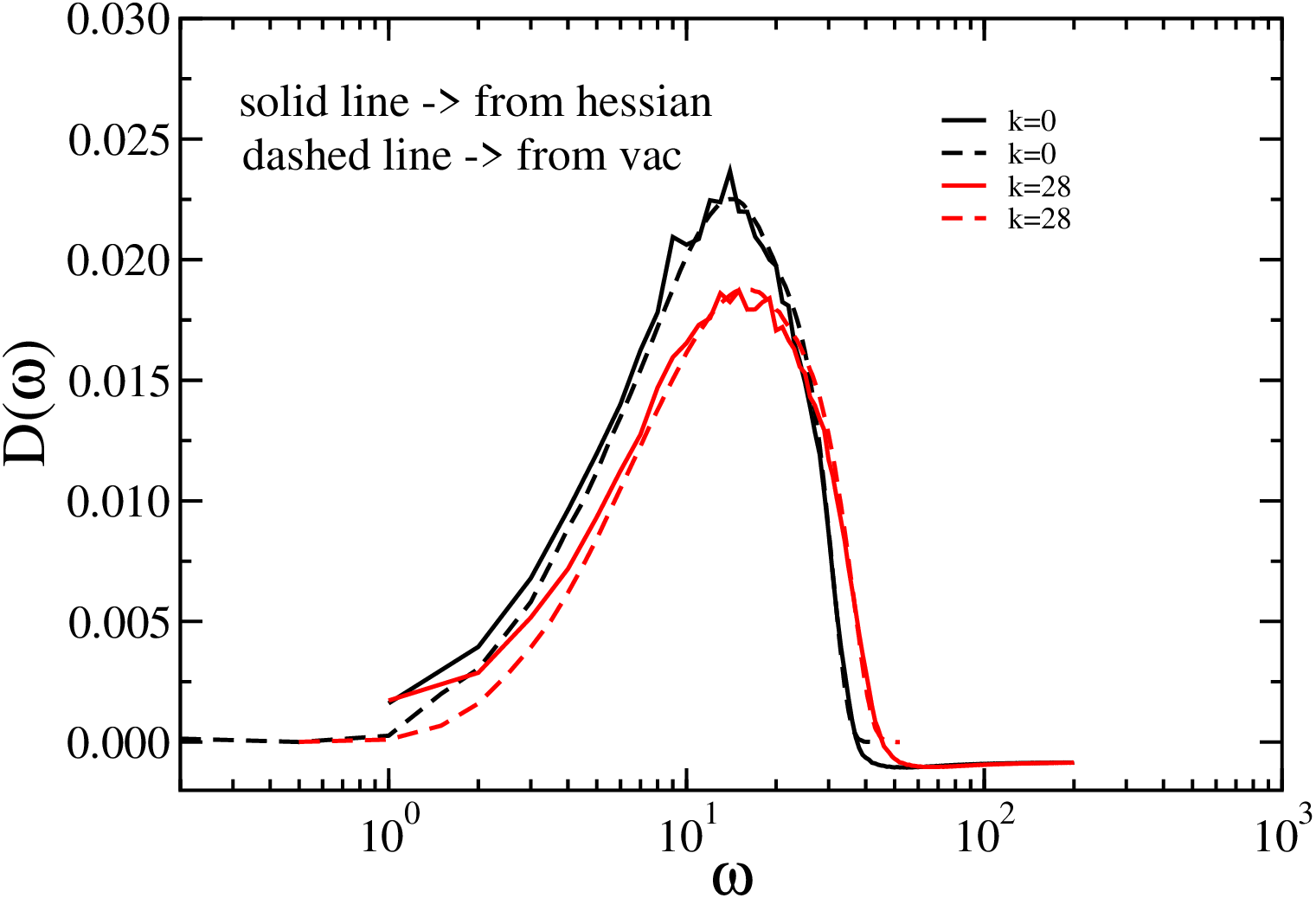}
\end{subfigure}
\caption{\emph{Density of states calculated from the Hessian and the velocity autocorrelation function for k=0 and k=28 systems. Both the methods show similar result. }}
\label{dos_compare}
\end{figure}

We have compared the density of states calculated from the calculation of Hessian and the Fourier transform of the velocity autocorrelation function. We find both the methods show a similar result in a density of states (see Fig.\ref{dos_compare}).

\textbf{Appendix II: Pinned system entropy }

In a pinned system, a fraction $c$ of the particles is pinned. The details about the pinned system have been discussed in simulation details (see section \ref{sec_details_pin}). 
Using the TI method the total entropy of moving particle in the pinned system, $S_{tot}$ is given by \cite{walter_original_pinning},
\begin{equation}
\begin{aligned}
  S_{tot} &= \frac{3M}{2} - \frac{3M}{2}\ln\Big( \frac{2\pi T}{h^{2}}\Big) + M(1-\ln\frac{N}{V}) \\
    & - \sum_{i=1}^{2}N_{i}\ln\frac{N_{i}}{N} + \beta^{*} \big<U\big>  -\int_{0}^{\beta^{*}}d\beta \big<U\big>
\end{aligned}   
    \label{S_total_pin_eq}
\end{equation}
\noindent where $N_{1}$ and $N_{2}$ are number of moving particles of type $A$ and $B$ respectively. $V$ is the total volume of system, $M$ is the total number of moving particles.
Total potential energy of system $U = U_{MP} + U_{MM}$, where $U_{MM}$ and $U_{MP}$ denotes the interaction energy between moving- moving particles and moving-pinned particles respectively.

\begin{figure}[!bth]
\begin{subfigure}[h]{0.4\textwidth}
\includegraphics[width=0.9\textwidth,trim = {0 0cm 0 0.0cm},clip]{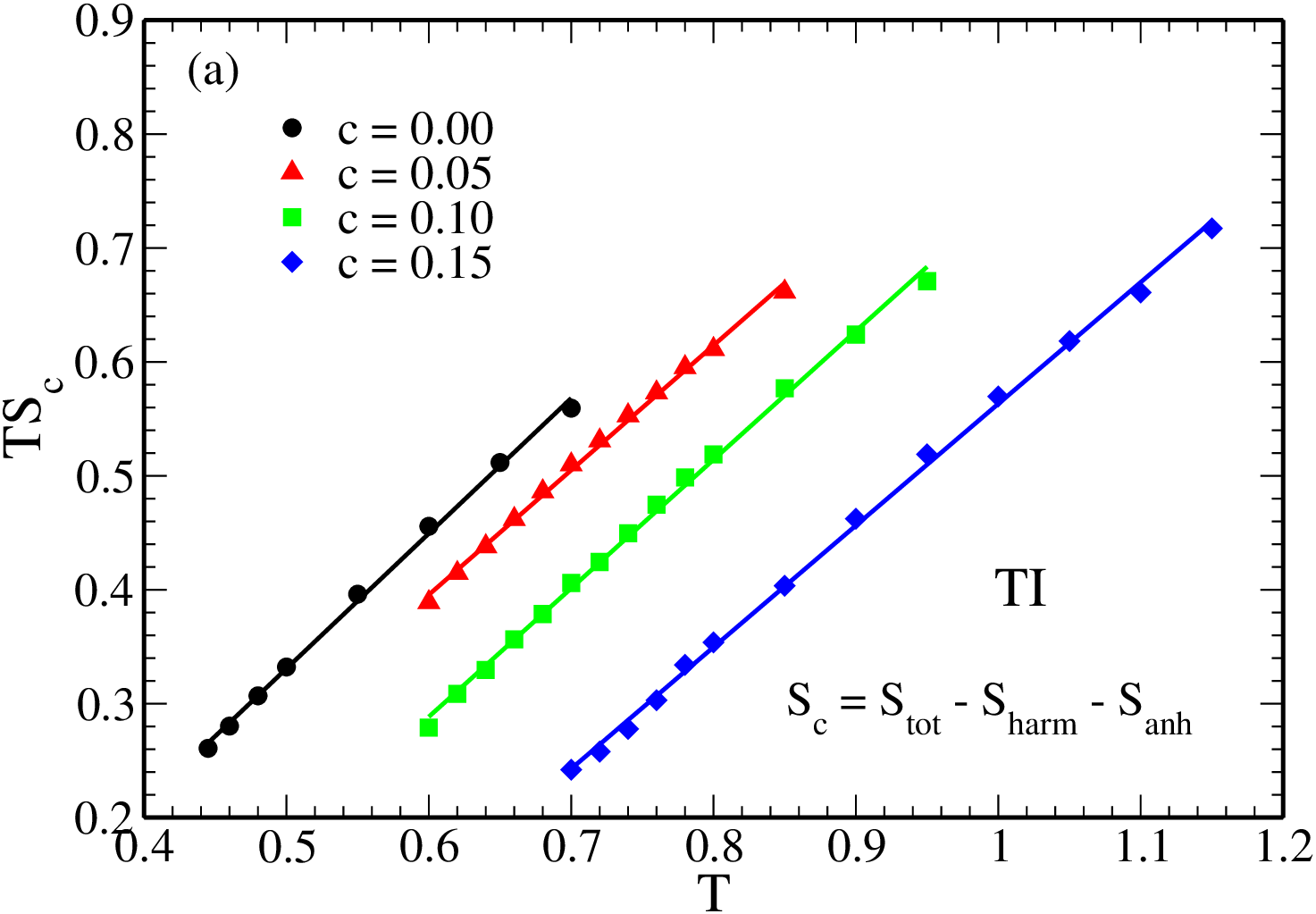}
\end{subfigure}
\begin{subfigure}[h]{0.4\textwidth}
\includegraphics[width=0.9\textwidth,trim = {0 0cm 0 0.0cm},clip]{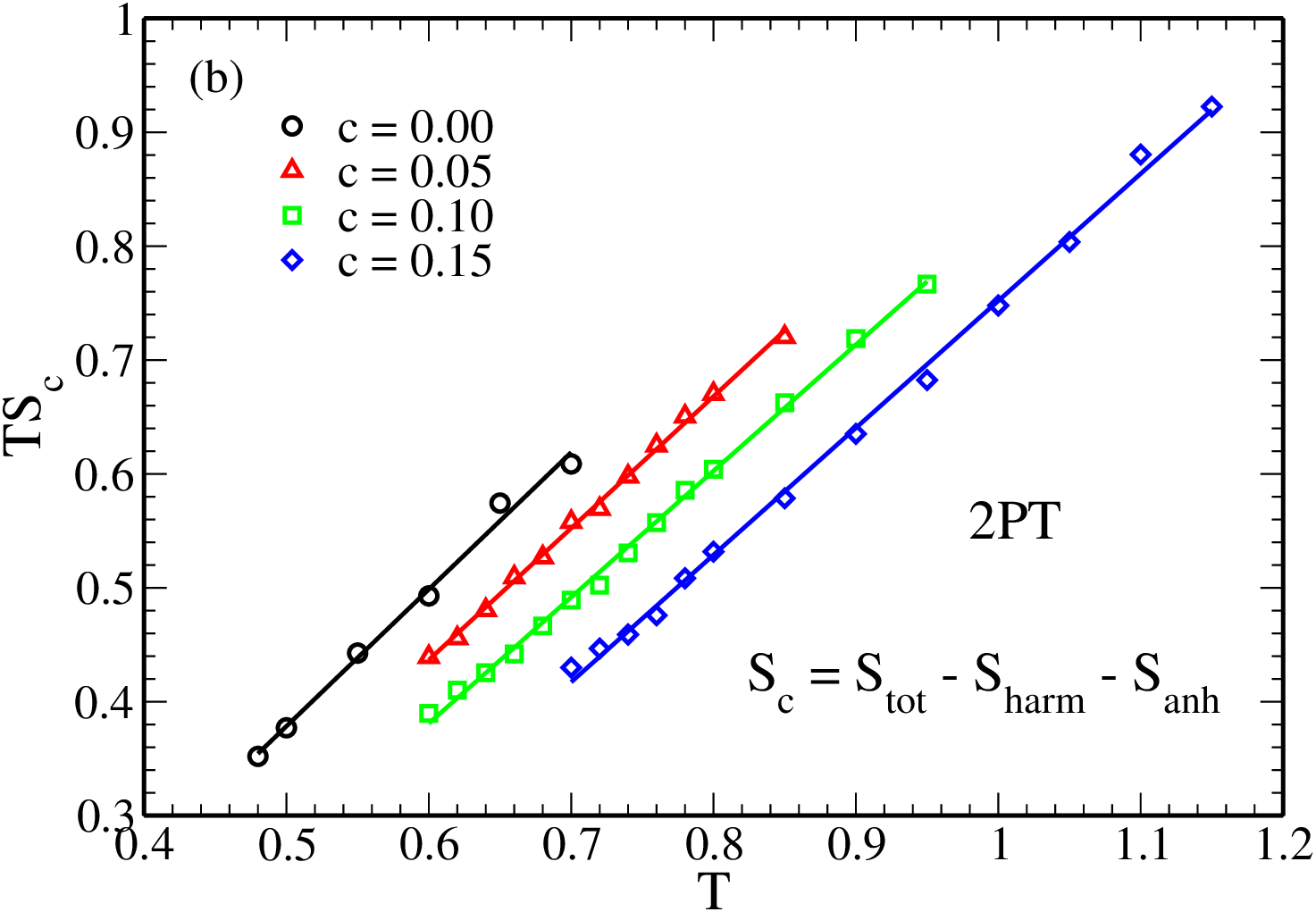}
\end{subfigure}
\caption{\emph{$TS_c \, vs. \, T$ for $c=0, 0.5, 0.10, 0.15$ systems using (a) the TI method and (b) the 2PT method. $S_c$ is computed by including the anharmonic contribution. $T^{TI}_{K}$ and $T^{2PT}_{K}$ increase with increasing pinning concentration but $T^{2PT}_{K}<T^{TI}_{K}$, see Table.\ref{table_compare_temp_pin_anh}.}}
\label{T_Sc_anh_diff_c_fig}
\end{figure}

\begin{figure}[!bth]
\begin{subfigure}[h]{0.4\textwidth}
\includegraphics[width=0.9\textwidth,trim = {0 0cm 0 0.0cm},clip]{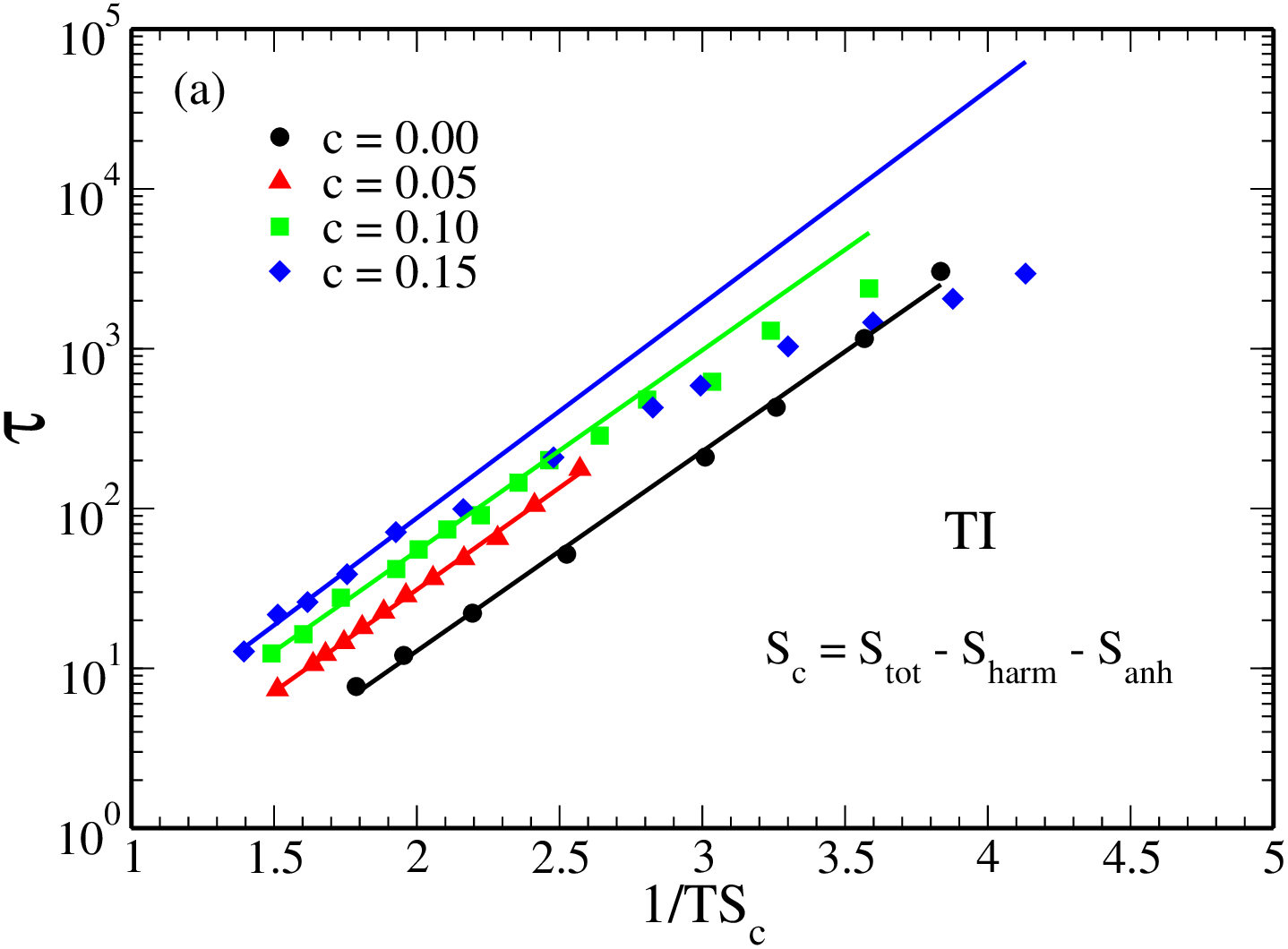}
\end{subfigure}
\begin{subfigure}[h]{0.4\textwidth}
\includegraphics[width=0.9\textwidth,trim = {0 0cm 0 0.0cm},clip]{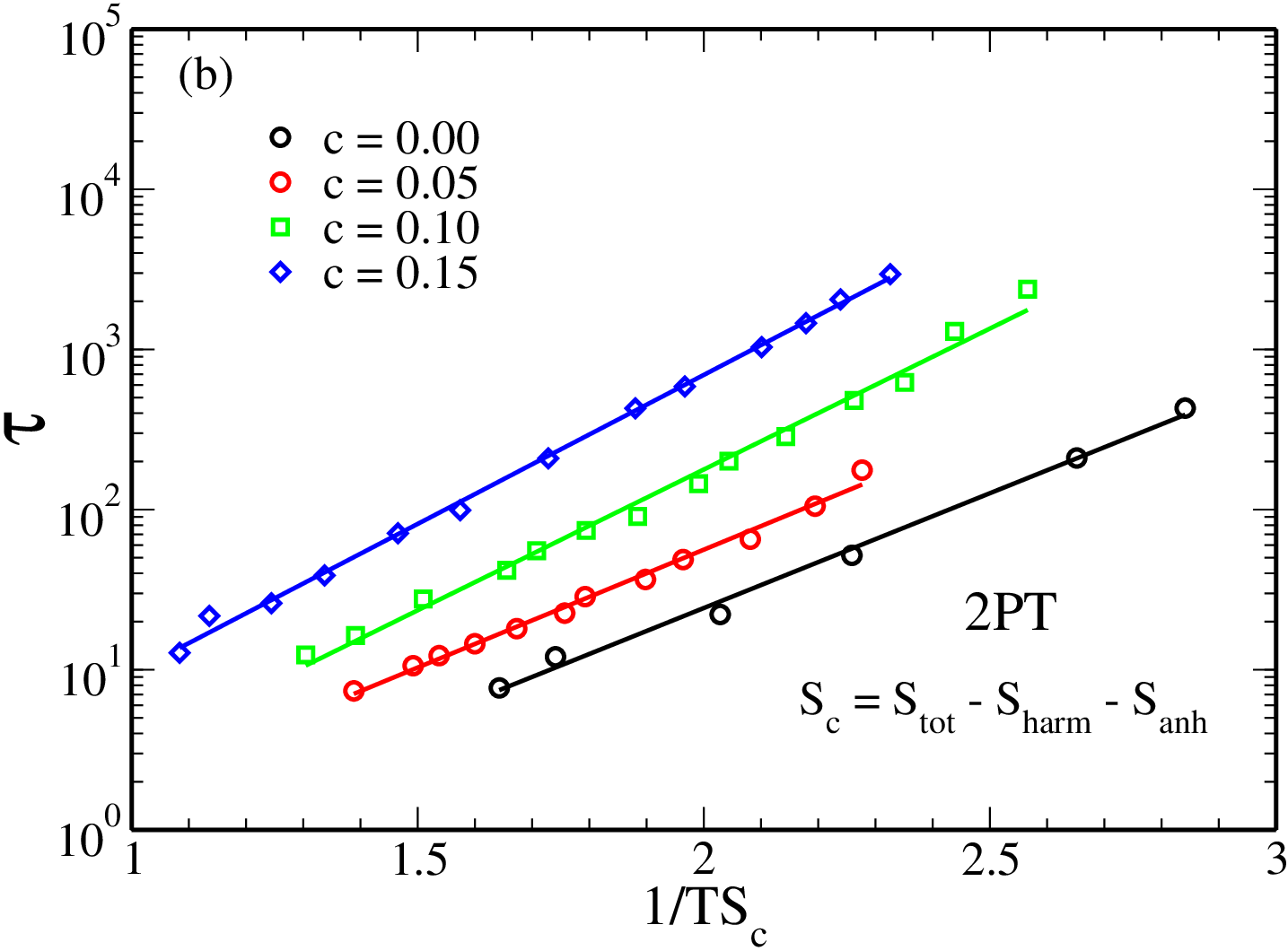}
\end{subfigure}
\caption{\emph{Testing the AG relation between $\tau$ and $\frac{1}{TS_{c}}$ for $c=0, 0.5, 0.10, 0.15$ systems using (a) the TI method and (b) the 2PT method. $S_c$ is computed by including the anharmonic contribution. In the temperature range studied here, the AG relation is violated when entropy is calculated using the TI method. However, the AG relation holds when entropy is calculated via the 2PT method. }}
\label{AG_anh_diff_c_fig}
\end{figure}

The temperature dependence of the configurational entropy after taking care of the anharmonic contribution is plotted in Fig.\ref{T_Sc_anh_diff_c_fig} (a) and the corresponding Adam-Gibbs plot is shown in Fig.\ref{AG_anh_diff_c_fig} (a). Even after the addition of the anharmonic contribution, the AG relationship is violated. In figure Fig.\ref{T_Sc_anh_diff_c_fig} (b) we plot the temperature dependence of the configurational entropy where the total entropy is calculated using the 2PT method and the anharmonic contribution is taken into consideration. We show the AG plot of the same data in Fig.\ref{AG_anh_diff_c_fig} (b). We find that when the total entropy is calculated using the 2PT method the AG relationship holds and also the temperature where the entropy vanishes is lower than that given by the TI method (see Table.\ref{table_compare_temp_pin_anh}.\\

\begin{table}
	\caption{The value of all characteristic temperatures for systems with different 'c' values. $T_{K}^{TI}$(anh),and $T_{K}^{2PT}$(anh) are Kauzmann temperature estimated from TI and 2PT respectively after addition of anharmonic approximation.}
	\begin{center}
	\addtolength{\tabcolsep}{+20.0pt}
		\begin{tabular}{ | l | l | l | }
			\hline
			 c  & T$_{K}^{TI}$ (anh) & T$_{K}^{2PT}$ (anh)  \\ \hline
			0.00  & 0.22 & 0.18      \\ \hline
			0.05  & 0.24 & 0.22      \\ \hline
		    0.10  & 0.34 & 0.26      \\ \hline
			0.15  & 0.47 & 0.33      \\ \hline

		\end{tabular}
		\label{table_compare_temp_pin_anh}
	\end{center}
\end{table}

{\bf Acknowledgment}\\
S.~M.~B thanks Walter Kob and Anshul D. S. Parmar for discussion, and SERB for funding. U.~K.~N., P.~P., and M.M. thanks CSIR, for the senior research fellowship. C.~D. acknowledges support from the Department of Science and Technology, Government of India.\\ 

{\bf Availability of Data}\\
The data that support the findings of this study are available from the corresponding author upon reasonable request.\\

{\bf References}\\

\end{document}